\documentclass[prd,preprint,aps,showpacs]{revtex4}
\usepackage{graphicx}
\usepackage{placeins}
\usepackage{lscape}
\usepackage{rotating}
\usepackage{float}
\usepackage{xcolor}
\definecolor{red}{rgb}{1,0,0}
\usepackage{graphics}

\newcommand{\newc}{\newcommand}
\newc{\gsim}{\lower.7ex\hbox{$\;\stackrel{\textstyle>}{\sim}\;$}}
\newc{\lsim}{\lower.7ex\hbox{$\;\stackrel{\textstyle<}{\sim}\;$}}
\newc{\gev}{\,{\rm GeV}}
\newc{\mev}{\,{\rm MeV}}
\newc{\ev}{\,{\rm eV}}
\newc{\kev}{\,{\rm keV}}
\newc{\tev}{\,{\rm TeV}}
\newc{\mz}{m_Z}
\newc{\mpl}{M_{Pl}}
\newc{\chifc}{\chi_{{}_{\!F\!C}}}
\newc\order{{\cal O}}
\newc\CO{\order}
\newc\CL{{\cal L}}
\newc\CY{{\cal Y}}
\newc\CH{{\cal H}}
\newc\CM{{\cal M}}
\newc\CF{{\cal F}}
\newc\CD{{\cal D}}
\newc\CN{{\cal N}}
\newc{\eps}{\epsilon}
\newc{\re}{\mbox{Re}\,}
\newc{\im}{\mbox{Im}\,}
\newc{\invpb}{\,\mbox{pb}^{-1}}
\newc{\invfb}{\,\mbox{fb}^{-1}}
\newc{\yddiag}{{\bf D}}
\newc{\yddiagd}{{\bf D^\dagger}}
\newc{\yudiag}{{\bf U}}
\newc{\yudiagd}{{\bf U^\dagger}}
\newc{\yd}{{\bf Y_D}}
\newc{\ydd}{{\bf Y_D^\dagger}}
\newc{\yu}{{\bf Y_U}}
\newc{\yud}{{\bf Y_U^\dagger}}
\newc{\ckm}{{\bf V}}
\newc{\ckmd}{{\bf V^\dagger}}
\newc{\ckmz}{{\bf V^0}}
\newc{\ckmzd}{{\bf V^{0\dagger}}}
\newc{\X}{{\bf X}}
\newc{\bbbar}{B^0-\bar B^0}

\newc{\sgn}{\mbox{sgn}\,}
\newc{\m}{{\bf m}}
\newc{\msusy}{M_{\rm SUSY}}
\newc{\munif}{M_{\rm unif}}
\newc{\slepton}{{\tilde\ell}}
\newc{\Slepton}{{\tilde L}}
\newc{\sneutrino}{{\tilde\nu}}
\newc{\selectron}{{\tilde e}}
\newc{\stau}{{\tilde\tau}}
\relax
\def\beq{\begin{equation}}
\def\eeq{\end{equation}}
\def\bea{\begin{eqnarray}}
\def\eea{\end{eqnarray}}
\newc{\ie}{{\it i.e.}}          \newc{\etal}{{\it et al.}}
\newc{\eg}{{\it e.g.}}          \newc{\etc}{{\it etc.}}
\newc{\cf}{{\it c.f.}}

\def\bino{{\tilde B}}

\def\beqn{\begin{eqnarray}}
\def\eeqn{\end{eqnarray}}
\relax
\newcommand{\ba}[1]{\begin{array}{#1}}
\def\ea{\end{array}}

\def\beq{\begin{equation}}
\def\eeq{\end{equation}}
\def\bea{\begin{array}}
\def\eea{\end{array}}

\def\to{\rightarrow}

\def\dis{\displaystyle}
\def\f{\frac}
\def\[{\left[}
\def\]{\right]}
\def\({\left(}
\def\){\right)}

\def\qs{{\sqrt{2}}}

\def\Ah{{\widehat{A}}}
\def\ms{{\widetilde{m}}}
\def\sm0{{\widetilde{m}_0}}

\def\cB{{\cos\beta}}
\def\tanB{{\tan\beta}}

\def\sM{{\widetilde{\cal M}}}

\def\sL{{\widetilde{L}}}
\def\sE{{\widetilde{E}}}

\def\tas{\tilde{\tau}}
\def\mus{\tilde{\mu}}
\def\slp{\tilde{l}}

\def\U1em{{U(1)_{\rm em}}}
\def\to{\rightarrow}

\def\CL{{\cal C}_L}

\def\sq2{\sqrt{2}}

\def\End{\end{document}}

\def\Dsl{\,\raise.15ex\hbox{/}\mkern-13.5mu D} 
\def\delsl{\raise.15ex\hbox{/}\kern-.57em\partial}
\def\Ksl{\hbox{/\kern-.6000em\rm K}}
\def\Asl{\hbox{/\kern-.6500em \rm A}}
\def\Qsl{\hbox{/\kern-.6000em\rm Q}}
\def\gradsl{\hbox{/\kern-.6500em$\nabla$}}
\def\bar#1{\overline{#1}}

\begin{document}
\draft
\title{A Numerical Analysis of the Supersymmetric Flavor Problem and Radiative
Fermion Masses}
\author{J.L. D\'{\i}az-Cruz$^{(a)}$, O. F\'elix Beltr\'an$^{(b)}$, M. G\'omez-Bock$^{(c)}$, R. Noriega-Papaqui$^{(d)}$, and A. Rosado$^{(e)}$}
\address{$^{(a)}$ Fac. Cs. F\'{\i}sico-Matem\'aticas, BUAP.
Apdo. Postal 1364, 72000 Puebla, Pue., M\'exico.\\
$^{(b)}$Fac. de Cs. de la Electr\'onica, BUAP. Apdo. Postal 1152, 72570  Puebla, Pue., M\'exico.\\
$^{(c)}$Instituto de F\'{\i}sica, UNAM. Apdo. Postal 20-364, 01000 M\'exico, D.F., M\'exico. \\
$^{(d)}$ Ctro. de Inv. en Matem\'aticas, UAEH. Carr. Pachuca-Tulancingo Km. 4.5, 42184 Pachuca, Hgo., M\'exico.\\
$^{(e)}$Instituto de F\'{\i}sica, BUAP. Apdo. Postal J-48, 72570 Puebla, Pue., M\'exico.}

\date{\today}
\begin{abstract}
We study the SUSY flavor problem in the MSSM, we are namely
interested in estimating the size of the SUSY flavor
problem and its dependence on the MSSM parameters. For that, we made
a numerical analysis randomly generating  the entries of the
sfermion mass matrices and then determinated which percentage of the
points are consistent with current bounds on the flavor violating
transitions on lepton flavor violating (LFV) decays $l_i \to l_j \, \gamma$. 
We applied two methods, mass-insertion approximation method (MIAM) and full diagonalization method (FDM).
Furthermore, we determined which fermion masses could be
radiatively generated  (through gaugino-sfermion loops) in a
natural way, using those random sfermion matrices. In general, the electron mass ge\-ne\-ra\-tion can be done with 30\% of
points for large $\tan\beta$, in both schemes the muon mass can be
generated by 40\% of points only when the most precise sfermion
splitting (from the FDM) is taken into
account.
\end{abstract}

\pacs{11.30.Hv, 11.30.Pb, 13.35.-r}
\keywords{Supersymmetric, flavor problem, MSSM, sfermion masses, LFV decays}
\maketitle

\setcounter{footnote}{0}

\setcounter{page}{2}

\setcounter{section}{0}

\setcounter{subsubsection}{0}
\hyphenation{ne-ce-ssa-ri-ly mi-ni-mal}

\narrowtext

\section{Introduction.}
Weak-scale supersymmetry (SUSY) \cite{review}, has notably become one of the
leading candidates for physics beyond the standard model,  by
supporting the mechanism of electroweak symmetry breaking (EWSB). Being a new
fundamental space-time symmetry, SUSY necessarily extends the SM particle
content by including superpartners for all fermions. Because the mass
spectrum of the superpartners needs to be lifted, SUSY must be softly broken,
this is needed so in order to maintain its ultraviolet properties. SUSY
breaking is parameterized in the Minimal Supersymmetric SM (MSSM) by the
soft-breaking lagrangian \cite{Chung:2003fi}; as an outcome, the combined effects of the
large top quark Yukawa coupling and the soft-breaking masses, make
 radiatively inducing  the breaking of the electroweak symmetry possible.
The Higgs sector of the MSSM includes two Higgs doublets, perhaps being the light
Higgs boson ($m_h \leq 125$ GeV)  the strongest prediction
of the model.

However, the soft breaking sector of the MSSM is often problematic
with low-energy flavor changing neutral currents (FCNC) without
making specific assumptions about its free parameters. Minimal
choices to satisfy those constraints, such as assuming
universality of squark masses, have been widely studied in
literature \cite{Chung:2003fi}. However, non-minimal flavor
structures could be generated in a variety of contexts. For
instance, within the context of realistic unification models
by the evolution of  soft-terms, from a high-energy GUT
scale to the weak scale. Similarly, models that attempt to
address the flavor problem, could induce sfermion soft-terms that
reflect the underlying flavor symmetry of the fermion sector
\cite{nondA,Mura}.

It is not a trivial task to find models of SUSY breaking that can
actually generate minimal and safe patterns. This is the so called
SUSY flavor problem. The known solutions include the following: {\it degeneracy}
\cite{gabbiani} (sfermions of different families have the same
mass), {\it proportionality} \cite{gabbiani} (trilinear terms are
proportional to the Yukawa terms), {\it decoupling}
\cite{Arkani-Hamed:1997ab} (superpartners are too heavy to affect
low energy physics) and {\it alignment} \cite{seiberg} (the same
physics that explains the pattern of fermion masses and mixing
angles, forces the sfermion mass matrices to be aligned with the
fermion ones, in a way that the fermion-sfermion-gaugino
vertices remain close to diagonal).

Sometimes the SUSY flavor problem is stated by saying that if the
sfermion mass matrix entries were randomly generated , most of these
points would lead to the exclusion of the MSSM. In this paper, we
would like to quantify the formerly statement, namely, we want to
estimate the {\it size} of the SUSY flavor
problem, and to determine its dependence on the parameters of the
MSSM. Then, one would like  to determine what would be left of the
SUSY flavor problem after Tevatron and LHC will deliver bounds on
the masses of the superpartners, or luckily a signal of their
presence! Rather, we focus on lepton sector, we particularly use the LFV decays $l_i \to l_j +\gamma$ to make
our point, namely to derive bounds on the parameters of the MSSM
and to determine the viability and interplay of the solutions above.

First, we evaluate the  LFV decays above using the mass-insertion
approximation method (MIAM), both for muon and tau decays. Our procedure will
consist first on writing the off-diagonal elements of the slepton
mass matrices as the product of $O(1)$ coefficients times an
average sfermion mass parameter, then we randomly generate  $10^5$
points for the $O(1)$ coefficients, and determine which fraction
of such points satisfies the current bounds on the LFV transitions. We
repeat this procedure for different values of other relevant
parameters of the MSSM, such as $\tan\beta$, gaugino masses,
$\mu$-parameter and the sfermion mass scale.

Next, to estimate how much  we  can trust the MIAM, we
compare those results with the ones coming from particular models
that enable to obtain exact diagonalization for the sfermion mass
matrices. Namely, we take into account that the constraints on
sfermion mixing coming from low-energy data, mainly suppress the
mixing between the first two family sleptons, but still allows large
flavor-mixings between the second- and third-family sleptons, {\it
i.e.} the smuon ($\mus$) and stau ($\tas$), which can be as large as
$O(1)$~\cite{FCNC}. Thus, we consider models where the mixing involving
the selectrons could be neglected, as it involves small off-diagonal
entries in the slepton mass matrices. But the $\mus-\tas$ mixing 
will involve large off-diagonal entries in the sfermion mass
matrices, which requires at least a partial diagonalization in order
to be treated in a consistent manner. Namely, in our models the
general $6\times 6$ slepton-mass-matrix will include a $4\times 4$
sub-matrix involving only the $\mus - \tas$ sector, which can be exactly
diagonalized , similarly to the squark case first discussed
in Ref.\cite{oursqmix}. Since we follow a bottom-up approach, we
simply take an {\it Ansatz} for the $A$-terms valid at the
TeV-scale; such large off-diagonal entries can be motivated by
considering the large mixing detected with atmospheric neutrinos
\cite{superkam}, especially in the framework of GUT models with
flavor symmetries. Then, we repeat the above method of random
generation for the parameters of the sfermion matrices, which
will then be  diagonalized. Armed with the exact expressions for the
mass and mixing matrices and the interaction lagrangian written in
terms of mass eigenstates, we evaluate the fraction of points that
satisfy all the LFV constraints coming from the  $\tau \to \mu
+ \gamma$ decays. The results with exact diagonalization for LFV tau decays
will be compared  with those obtained using the MIAM \footnote{Recently a similar analysis was presented in
Ref.\cite{Paradisi:2005fk}.}.

Another aspect of the Flavor Problem involves the possibility to
radiatively induce  the fermion masses which are known to be possible
within SUSY through sfermion-gaugino loops. Here, we shall determine
which fraction of points can generate correctly the fermion masses
through sfermion-gaugino loops. Again, we are interested in
comparing the results obtained using FDM with
those of the MIAM. Implications for LFV in
the Higgs sector are discussed in Refs.
\cite{Diaz-Cruz:2002er,Diaz-Cruz:1999xe}.

The organization of this paper goes as follows: in section II we
discuss the SUSY flavor problem in the lepton sector, using the
mass-insertion approximation. This section includes the evaluation
of the radiative LFV loop transitions ($l_i \to l_j +\gamma$)
with a random generation of the slepton $A$-terms. Then, in section
III we present an {\it Ansatz} for soft breaking trilinear terms,
the diagonalization of the resulting sfermion mass matrices, and
we repeat the calculus of the previous section. The radiative
generation of fermion masses is discussed in detail in section IV,
within the context of the MSSM. Finally, our conclusions are
presented in section V.
\section{The SUSY flavor problem in the Super CKM basis.}

\subsection{The slepton mass matrices in the MSSM}

First, we discuss the slepton mass matrices and the gaugino-lepton-slepton interactions. The MSSM
soft-breaking slepton sector contains the following quadratic
mass-terms and trilinear $A$-terms: 
\beq \bea{l}
{\cal L}_{soft}=-\sL_i^\dag (M_{\sL}^{2})_{ij}\sE_j
-\sE_i^\dag (M_{\sE}^{2})_{ij}\sE_j  
+( A_{l}^{ij}\sL_i H_d\sE_j  + {\rm h.c.} )\,, \eea
\label{eq:A-term} 
\eeq
where $\sL_i$ and ${\tilde{E_j}}$ denote the doublet and singlet slepton
fields, respectively, with $i,j(=1,2,3)$ being the family indices.
For the charged slepton sector, this gives a generic $6\times6$
mass matrix given by
\beq \sM^2_l =\left\lgroup
         \bea{ll}
          M_{LL}^2         &  M_{LR}^2\\[1.5mm]
          M_{LR}^{2\,\dag}   &  M_{RR}^2
         \eea
         \right\rgroup ,
\label{eq:MU6x6}
\eeq
where
\beq
\bea{ll}\label{eq:MU3x3}
M_{LL}^2 &= M_{\sL}^2+M_l^2+\f{1}{2}\cos2\beta \,(2m_W^2-m_Z^2)\,, \\[2mm]
M_{RR}^2 &= M_{\sE}^2+M_l^2-\cos2\beta\sin^2\theta_W\, m_Z^2\,, \\[1mm]
M_{LR}^2 &= \dis A_l v\,\cB/\sqrt{2}-M_l\,\mu\,\tanB \,. \eea
\eeq
Here $m_{W,Z}$ denote the $W^\pm$ and $Z^0$ masses
and $M_l$ being the lepton mass matrix (for convenience, we will choose
a basis where $M_l(=M_l^{\rm diag})$ is diagonal).

In our {\it minimal} scheme, we consider all large LFV that {\it
solely} come from the non-diagonal entries of the $A_l$-terms in
the slepton-sector, such that respects the low-energy
constrains and CCB-VS bounds \cite{CCBVS}. In the Super CKM basis,
the gaugino-slepton-lepton interactions are diagonal in flavor
space, while flavor-violation associated with the off-diagonal
entries of the slepton mass matrices are treated as
perturbations, {\it i.e.}, mass-insertions. We shall write the
off-diagonal soft-terms as
 \beq
( M^2_{MN})_{off-diag}=
z^l_{MN} \cdot \sm0^2 ,
\eeq
where $M,N: \, L,R$, $\sm0$ denotes an average slepton mass
scale and the coefficients $z^l_{MN}$ will be taken as random
coefficients of $O(1)$.

\subsection{Bounds on the soft-breaking parameters from the LFV
decay $l_i \to l_j \, \gamma$}

Here, we are interested in obtaining bounds on the 
$z^l_{MN}$ and $\sm0$ parameters, applying the MIAM in order to
evaluate the LFV transition $\mu \to e +\gamma$ and $\tau \to \mu
(e) +\gamma$. Within this method, the expression for the branching
ratio $BR(l_i \to l_j +\gamma)$, including the photino
contributions, can be written as follows \cite{gabbiani}:
\begin{eqnarray}\label{BRij1}
BR(l_i \to l_j \, \gamma) &=&\f{\alpha^3}{G^2_{F}} \f{12
\pi}{m^4_{\tilde{l}}} \left \{ \left|
M_3(x_{\tilde{\gamma}})(\delta_{ij}^l)_{LL}+\f{m_{\tilde{\gamma}}}{m_{l_i}}
M_1(x_{\tilde{\gamma}})(\delta_{ij}^l)_{LR} \right|^2 + (L
\leftrightarrow R) \right \}
\nonumber\\
& & \times BR(l_i \to l_j \, \nu_i \, \bar{\nu}_j),
\end{eqnarray}
where $M_1$ and $M_3$ are the loop functions, which are given below; $(\delta_{ij}^l)_{MN}=\widetilde{M}^2_{MN}/\sm0^2$ and 
$ x_{\tilde{\gamma}} \equiv(m_{\tilde{\gamma}}/\sm0)^2$.

Assuming that  the $(\delta_{ij}^l)_{LR}$ term exclusively 
contributes to the branching ratio, and considering 
\beq
\left( \widetilde{M}^2_{LR} \right)_{ij} = \frac{v_1}{\sqrt{2}}(A^l_{LR})_{ij}, \qquad i \neq j.
\eeq
with $v_1=v \cos\beta$ and $(A^l_{LR})_{ij}=(z^l_{LR})_{ij}
\cdot \sm0$, we obtain the following expression for $(\delta_{ij}^l)_{LR}$
\beq \label{eq:delta}
(\delta_{ij}^l)_{LR} = \frac{\left( \widetilde{M}^2_{LR}
\right)_{ij}}{\sm0^2} = \frac{\cos\beta}{\sqrt{2}} \,
\frac{v}{\sm0} \cdot (z^l_{LR})_{ij}.
\eeq
Finally, replacing the above expression in
Eq. (\ref{BRij1}), we obtain the following expression:
\beq BR(l_i \to l_j \, \gamma)  \approx \f{\alpha^3}{G^2_{F}} \f{6
\pi} {m^4_{\tilde{l}}} \left( \f{m_{\tilde{\gamma}}}{m_{l_i}}
\right)^2 \left| M_1(x_{\tilde{\gamma}}) \right|^2 \, \cos^2\beta
\, \left( \frac{v}{\sm0} \right)^2 \cdot (z^l_{LR})^2_{ij} \cdot
BR(l_i \to l_j \, \nu_i \, \bar{\nu}_j), 
\eeq
where $m_{{\tilde{l}}_i} \approx \sm0ÿ$ and
\beq
M_1(x_{\tilde{\gamma}})=\frac{1 + 4 x - 5 x^2 + 4 x \ln (x) + 2 x^2 \ln (x)}
{2 (1 - x)^4}.
\eeq
In order to discuss the processes $\mu \to e \,\gamma$ and $\tau \to \mu \, \gamma (e \, \gamma)$, we
shall make use of the following experimental results: $BR(\mu
\to e \, \nu_{\mu} \, \bar{\nu}_{e}) \approx 100 \% $;
$BR(\tau \to \mu \, \nu_{\tau} \, \bar{\nu}_{\mu}) \approx
17.36 \% $; $BR(\tau \to e \, \nu_{\tau} \, \bar{\nu}_{e})
\approx 17.84 \% $, respectively \cite{partdata}.

Then, we calculate the bino contributions to $BR(l_i \to l_j \,
\gamma)$ following Ref.\cite{Paradisi:2005fk} and  obtain
\beq\label{BRij1Pa} 
BR(l_i \to l_j \, \gamma)  \approx \f{25
\pi}{3\cos^4\theta_W} \f{\alpha^3}{G^2_{F}}
\f{\widetilde{m}^4}{m^8_{L}} \left( \f{m_{1}}{m_{l_i}} \right)^2
\left \{ \left| M_1(a_L)(\delta_{ij}^l)_{LR} \right|^2 \right \} \,
BR(l_i \to l_j \, \nu_i \, \bar{\nu}_j), 
\eeq 
where $a_L=m_{1}^2/m^2_{L}$, $m_{1}(m_{U})=m_{1/2}$ is the
gaugino mass (in this case the mass of the $\bino$), and
$m^2_{L}(m_{U})=m^2_{0}$ is a common scalar mass.

If we consider the approximation $m^2_{L}=m^2_{0}=\widetilde{m}^2_{0}$ and $m_{1}=m_{\bino}$, then
Eq.(\ref{BRij1Pa}) reduces to
\beq\label{BRij1Pa2} 
BR(l_i \to l_j \, \gamma)  \approx \f{25
\pi}{3\cos^4\theta_W} \f{\alpha^3}{G^2_{F}}
\f{1}{\widetilde{m}^4_{0}} \left( \f{m_{\bino}}{m_{l_i}} \right)^2
\left \{ \left| M_1(x_{\bino})(\delta_{ij}^l)_{LR} \right|^2
\right \} \, BR(l_i \to l_j \, \nu_i \, \bar{\nu}_j) ,
\eeq
where $x_{\bino} \equiv (m_{\bino}/\widetilde{m}_{0})^2$.

Now, our numerical analysis is based on a random generation of
the parameters $(z^l_{LR})_{ij}$  ($10^5$ points are generated)
and then studying their effects on the LFV transitions. Our
results for $\mu \to e \, \gamma$ are shown in Fig. 1, assuming
$\tan\beta=15$ for $x_{\bino}=0.3,1.5,5$.

Fig. 1 illustrates the severity of the SUSY flavor
problem for low sfermion masses. One can see that even for
$\sm0=1$ TeV almost $100 \%$ of the randomly generated points
are experimentally excluded, while one needs to have $\sm0 \approx
10$ TeV in order to obtain that approximately $10 \%$ of the generated
points satisfy the current bound on $\mu \to e \, \gamma$. On other hand, 
larger gaugino masses help to ameliorate the problem, but not
much. For instance, assuming $x_{\bino}=5$ and $\tan\beta=15$,
implies that even for $\sm0=10$ TeV, the percentage of
acceptable points only raises up to  $18 \%$.

Current bounds on tau decays do not pose such severe problem, as
is shown in Figs. 2. In this case most of the randomly generated
points satisfy the bounds on $\tau \to \mu \, \gamma$ and $\tau
\to e \, \gamma$.
For instance, in the case of $\tau \to \mu \, \gamma$, with $x=0.3$ and
$\tan\beta = 15$ (see Fig. 2(a)); it is obtained that for $m_{\widetilde l} = 200$
GeV approximately $10 \%$ of the points are accepted by experimental data.
However, this percentage increases with the slepton mass, and  for
$m_{\widetilde l} \geq 400$ GeV about $100 \%$ of the points are accepted by
experimental data. In Fig. 2(b),
we notice that a similar behavior is obtained for $\tau \to e \, \gamma$.
We can also notice in Fig. 2(a) (Fig. 2(b)) that for $x=0.5$ and $\tan\beta = 15$
in the case $\tau \to \mu \, \gamma$ ($\tau \to e \, \gamma$)
requires slepton masses, under $m_{\widetilde l} \geq 220$ GeV
($m_{\widetilde l} \geq 180$ GeV) in order to get $100\%$ of the points as
acceptable by experimental data.

\section{The SUSY flavor problem beyond the mass-insertion approximation.}

Now, we shall consider SUSY FCNC schemes where the general
$6\times 6$ slepton-mass-matrix reduces down to a $4\times 4$
matrix involving only the $\mus - \tas$ sector, similarly to the
quark sector discussed in Ref. \cite{oursqmix}. In this case, $\mus-\tas$ 
flavor-mixings  can be as large as $O(1)$. Although such large mixing could be related
to the large $\nu_\mu -\nu_\tau$ mixing observed in atmospheric
neutrinos \cite{superkam}, we shall follow a bottom-up approach,
where we simply take as an {\it Ansatz} the following form of the
$A$-terms, taken also to be real and valid at the TeV-scale.
Here, we consider two {\it Ansatz} kinds for $A$-terms, which are used for the  
diagonalization of fermion mass matrices.

\subsection{Diagonalization of fermion mass matrices}
 
\underline{{1. {\it Ansatz} A}}

The reduction of the slepton mass matrix proceeds, for instance,
by considering at the weak scale the following $A$-term ({\it
Ansatz} A): \beq A_l =
      \left\lgroup
      \bea{ccc}
      0 & 0 & 0\\
      0 & 0 & z\\
      0 & y & 1
      \eea
      \right\rgroup A_0 \,,
\label{eq:Alep}
\eeq
where $y$ and $z$ can be of $O(1)$, representing a naturally large
flavor-mixing in the $\mus - \tas$ sector. Actually, the zero entries could
be of $O(\epsilon)$, with $\epsilon \ll 1$, and their effect could be
treated using the MIAM. Moreover, if we identify the
non-diagonal $A_l$ as the only source of the observable LFV phenomena, this would
implies that the slepton-mass-matrices $M_{\sL,\sE}^2$ in
Eqs.\,(\ref{eq:MU6x6})-(\ref{eq:MU3x3}) to be nearly diagonal.
For simplicity, we define
\beq
M_{LL}^2 \,\simeq\, M_{RR}^2\, \simeq\,\sm0^2\,{\bf I}_{3\times3}\,,
\label{eq:Degen}
\eeq
with $\sm0$ being a common scale for scalar-masses.

Within this minimal scheme, we observe that the first  slepton family
$\tilde{e}_{L,R}$ decouples from the rest in (\ref{eq:MU6x6})
so that, in the slepton basis $(\mus_L,\,\mus_R,\,\tas_L,\,\tas_R)$,
the $6\times6$ mass-matrix is reduced to the following
$4\times4$ matrix,
\beq
\sM_{\slp}^2  =
      \left\lgroup
      \bea{cccc}
      ~{\ms}_{0}^2~  &   0             &   0            &   A_z\\[1mm]
      0              &   ~\ms_{0}^2~   &   A_y~         &   0  \\[1mm]
      0              &   A_y~          &  ~\ms_{0}^2~   &   X_\tau\\[1mm]
      A_z~           &   0             &   X_\tau~         &  ~\ms_{0}^2
      \eea
      \right\rgroup
\label{eq:Mct4x4}
\eeq
where
\beq
\bea{l}
A_y = y\Ah\,,~~~A_z = z\Ah\,,~~~\Ah = Av\,\cB/\sqrt{2}\,,~~~  
X_\tau = \Ah - \mu\,m_\tau\,\tanB \,.                     \\ 
\eea
\label{eq:MctADD}
\eeq
The reduced slepton mass matrix (\ref{eq:Mct4x4}) allows an exact diagonalization. Therefore,
when evaluating loop amplitudes one can use the exact slepton
mass-diagonalization and compare the results with those obtained from the
popular but crude MIAM.

We now have obtained the mass-eigenvalues of the eigenstates 
$(\mus_{1},\,\mus_{2},\,\tas_1,\,\tas_2)$ for any $(y,\,z)$, given as:
\beq \bea{ll}
M_{\mus_{1,2}}^2 & =\sm0^2 \mp\f{1}{2}|\sqrt{\omega_+}-\sqrt{\omega_-}|\,, \\[2mm]
M_{\tas_{1,2}}^2 & = \sm0^2
\mp\f{1}{2}|\sqrt{\omega_+}+\sqrt{\omega_-}|\,, \eea \label{eq:Mass}
\eeq 
where $~\omega_\pm = X_\tau^2+(A_y\pm A_z)^2\,$. From
(\ref{eq:Mass}), we can deduce the mass-spectrum of the $\mus - \tas$ sector
 as $ M_{\tas_{1}} < M_{\mus_{1}} < M_{\mus_{2}} <
M_{\tas_{2}}$. 

The $4\times4$ rotation
matrix of the diagonalization is given by, \beq \bea{l} \left\lgroup
\bea{l}
\mus_L\\
\mus_R\\
\tas_L\\
\tas_R
\eea
\right\rgroup
\!\!=\!\!
\left\lgroup
\bea{rrrr}
 c_1c_3  &  c_1s_3  & s_1s_4  &  s_1c_4  \\
-c_2s_3  &  c_2c_3  & s_2c_4  & -s_2s_4  \\
-s_1c_3  & -s_1s_3  & c_1s_4  &  c_1c_4  \\
 s_2s_3  & -s_2c_3  & c_2c_4  & -c_2s_4
\eea
\right\rgroup
\!\!
\left\lgroup
\bea{l}
\mus_1\\
\mus_2\\
\tas_1\\
\tas_2
\eea
\right\rgroup , 
\eea
\label{eq:rotation}
\eeq
with
\beq ~~~
s_{1,2}=\dis \f{1}{\qs}
\[1-\f{X_{\tau}^2\mp A_y^2\pm A_z^2}{\sqrt{~\omega_+\omega_-}} \]^{1/2}\!,~~~
s_4
=\f{1}{\qs}\,,
\label{eq:rotation2}
\eeq
and $s_3=0\,(1/\qs)$ if $yz=0$($yz \neq 0$).

In Fig.3, we plot the slepton spectra as functions of $z$ for $\sm0=100,500$ GeV
and $\sm0=1,10$ TeV, taking
$\tan\beta=15$. We can observe that both $\tas_1$ and
$\tas_2$ differ significantly from the common scalar mass $\sm0$;
stau $\tas_1$ can be as light as about
$100-300$ GeV, which  have an important effect in the loop
calculations. Furthermore, even for $z \simeq 0.5$ the smuon
masses can differ from $\sm0$ for 30-50 GeV. With these
mass values the slepton phenomenology would have to be
reconsidered, since one is not allowed to sum over all the
selectrons and smuons, for instance, when evaluating slepton
cross-sections, as it is usually assumed in the constrained MSSM.
We can also observe in Fig. 3 that $m_{\mus_1} - m_{\tas_1}$
and $m_{\mus_2} - m_{\tas_2}$ almost behave  constant as one varies
the parameter $z$ in the range
$0 \leq z \leq 1$. However, the differences $m_{\mus_2} -
m_{\mus_1}$ and $m_{\tas_2} - m_{\tas_1}$ are sensitive to the
non-minimal flavor structure. Besides, such splitting will affect
the results for LFV transitions and the radiative fermion mass
generation.

\bigskip
 
\underline{{2. {\it Ansatz} B}}

Now, we will reduce the slepton mass matrix by
considering another $A$-term at the weak scale ({\it Ansatz}
B):
\beq A_l =
      \left\lgroup
      \bea{ccc}
      0 & 0 & 0\\
      0 & w & y\\
      0 & y & 1
      \eea
      \right\rgroup A_0 \,,
\label{eq:Alepb} \eeq where $w$ and $y$ can be of $O(1)$, and as {\it Ansatz}
A, the zero entries could be of $O(\epsilon)$, with
$\epsilon \ll 1$.  For this case, we take the same considerations of {\it Ansatz}
A.  Again,  the first 
slepton  family $\tilde{e}_{L,R}$ decouples from the rest in
(\ref{eq:MU6x6}) and we obtain
 \beq \sM_{\slp}^2  =
      \left\lgroup
      \bea{cccc}
      ~{\ms}_{0}^2~  &   A_w~          &   0            &   A_y\\[1mm]
      A_w~           &   ~\ms_{0}^2~   &   A_y~         &   0  \\[1mm]
      0              &   A_y~          &  ~\ms_{0}^2~   &   X_\tau\\[1mm]
      A_y~           &   0             &   X_\tau~         &  ~\ms_{0}^2
      \eea
      \right\rgroup
\label{eq:Mct4x4b} 
\eeq 
Here, $A_w = w\Ah$, and $A_y$, $\Ah$ and $X_\tau$ are the same of Eq (\ref{eq:MctADD}).

For this case, mass-eigenvalues of the eigenstates
$(\mus_1,\,\mus_2,\,\tas_1,\,\tas_2)$ for any $(w,\,y)$ have the following expressions:
\beq \bea{ll}
M_{\mus_{1,2}}^2 & =  \frac{1}{2}(2 \sm0^2 \pm A_w \pm X_{\tau} \mp R),\nonumber\\
M_{\tas_{1,2}}^2 & =  \frac{1}{2}(2 \sm0^2\mp A_w \mp X_{\tau}\mp R),
\eea\label{eq:Massb} 
\eeq 
where
$R=\sqrt{4A_y^2+\left(A_w-X_{\tau} \right)^2}$. From
(\ref{eq:Massb}) and considering $\mu < 0$, the
mass-spectrum of the $\mus - \tas$ sector as $ M_{\tas_{1}} <
M_{\mus_{1}} < M_{\mus_{2}} < M_{\tas_{2}}$.

With this ansatz, the slepton spectra as functions of $y$ for
$\sm0=100,500$ GeV, $\sm0=1,10$ TeV with
$\tan\beta=15$, by considering
$w=0.0,\,0.5,\,1.0$ have a similar behavior as in the case of {\it
Ansatz} A.

By defining 
\beq\label{defphi}
\sin\phi=\frac{2A_y}{\sqrt{4A_y^2+(A_w-X_{\tau})^2}} \, ,
\mbox{\hspace{1cm}}\
\cos\phi=\frac{2A_w-X_{\tau}}{\sqrt{4A_y^2+(A_w-X_{\tau})^2}} \, ,
\eeq \noindent the $4\times4$ rotation matrix of the diagonalization
is given by, \beq \bea{l} \left\lgroup \bea{l}
\mus_L\\
\mus_R\\
\tas_L\\
\tas_R \eea \right\rgroup \!\!=\frac{1}{\sqrt{2}}\!\! \left\lgroup
\bea{rrrr}
-s_{\xi} & s_{\xi}  & -c_{\xi} & c_{\xi} \\
-s_{\xi} & -s_{\xi} & c_{\xi}  & c_{\xi} \\
c_{\xi}  & -c_{\xi} &-s_{\xi}  & s_{\xi}  \\
c_{\xi}  & c_{\xi}  & s_{\xi}  & s_{\xi} \eea
\right\rgroup \!\! \left\lgroup \bea{l}
\mus_1\\
\mus_2\\
\tas_1\\
\tas_2 \eea
\right\rgroup , 
\eea \label{eq:rotationb} 
\eeq
where  $s_{\xi}\equiv \sin(\phi/2)$ and $c_{\xi}\equiv
\cos(\phi/2)$. 

\subsection{Gaugino-sfermion interactions}
The interaction between gauginos and lepton-slepton pairs can be
written as follows:
\beq
{\cal{L}}_{int}= \bar{\chi}^0_m
   [ \eta^{mL}_{\alpha k} P_L + \eta^{mR}_{\alpha k} P_R]
          \slp_\alpha  l_k +h.c.,
\eeq 
where $\chi^0_m$ ($m=1,...,4$) denotes the neutralinos, while
$\slp_{\alpha}$ correspond to the mass-eigenstate sleptons. The
factors $\eta^{mN}_{\alpha k}$ are obtained after substituting
the rotation matrices for both  neutralinos and sleptons in the
interaction lagrangian.

To carry out the forthcoming analysis of LFV transitions,
we choose to work with the simplified case $y=z$,
which gives: $c_1=c_2=c_{\slp}$, $s_1=s_2=s_{\slp}$ and
$c_3=s_3=c_4=s_4=\f{1}{\sq2}$.
 The expressions for $\eta^{mL,R}_{\alpha k}$
simplify further when the neutralino is taken
as the bino, which we will assume in the calculation of Higgs
LFV decays; the resulting coefficients ($\eta^{L,R}_{\alpha k}$)
are shown in Table I.
\begin{table}

\begin{center}
\begin{tabular}{|c|c|c|c|c|c|c|c|c|}
\hline\hline
$({\slp}_{\alpha},l_k)$ & $(\mus_1,\mu)$  &  $(\mus_1,\tau)$  &
$(\mus_2,\mu)$ &  $(\mus_2,\tau)$  & $(\tas_1,\mu)$ &
$(\tas_1,\tau)$  & $(\tas_2,\mu)$  & $(\tas_2,\tau)$ \\
\hline
$\eta^{L}_{\alpha k}$   &  $-c_{\slp} \f{g_1}{2}$ &  $s_{\slp} \f{g_1}{2}$ &
$-c_{\slp} \f{g_1}{2}$ &  $s_{\slp} \f{g_1}{2}$ &  $-s_{\slp} \f{g_1}{2}$ &
$-c_{\slp} \f{g_1}{2}$ &  $-s_{\slp} \f{g_1}{2}$ &  $-c_{\slp} \f{g_1}{2}$ \\
\hline
$\eta^{R}_{\alpha k}$  &  $-c_{\slp} g_1$ &  $s_{\slp} g_1$ &
$-c_{\slp} g_1$ &  $-s_{\slp} g_1$ &  $s_{\slp} g_1$ &
$c_{\slp} g_1$&  $-s_{\slp} g_1$&  $-c_{\slp} g_1$ \\
\hline\hline
\end{tabular}
\end{center}
\caption{Slepton-lepton-neutralino couplings ($\eta^{mN}_{\alpha k}$)
for the case when $y=z$ and $\chi^0_1=\tilde{B}$. }
\label{tab:param}
\end{table}
\bigskip

\subsection{Bounds on the LFV parameters from $\tau \to \mu + \gamma$}

Here, we are interested in determining which fraction of points in
parameter space satisfy current bounds on LFV tau decays, when the
exact slepton mass-diagonalization is applied; again we generate
$10^5$ random values of $O(1)$ for the parameter $z$ appearing in
the soft-terms, and fix the values of $\widetilde{m}_0$,
$\widetilde{M}$ and $\tan\beta$. Using interaction
lagrangian (21) the one can write the general
expressions for the SUSY contributions to the decays $\tau \to \mu
+ \gamma$ given in Ref. \cite{hisanoetal}. The expression for  $\Gamma(\tau
\to \mu + \gamma)$, including the $ \mus$ and $\tas$ contributions, is written as follows: 
\beq
 \Gamma (\tau \to \mu + \gamma)= \f{\alpha  m^5_{\tau}}{4\pi}
   [ \sum_\alpha |A_{L\alpha}|^2+|A_{R\alpha}|^2 ],
\eeq
where
\beq
 A_{R\alpha}  = \f{1}{ 32\pi^2 m^2_{{\slp}_\alpha} }
  [\eta^R_{{\slp}_\alpha \tau} \eta^R_{{\slp}_\alpha \mu} f_1(x_\alpha)
          + \eta^R_{{slp}_\alpha \tau} \eta^L_{{\slp}_\alpha \mu}
       \f{m_{\tilde{B}} }{m_\tau} f_2(x_\alpha) ],
\eeq 
with $x_\alpha=m^2_{\tilde{B}}/m^2_{{\slp}_{\alpha} }$, and
the functions $f_{1,2}(x_\alpha)$ are given in Ref.
\cite{hisanoetal}. $A_{L\alpha}$ is obtained by making the
substitutions $L,R \to R,L$ in Eq.(23). The expressions for the
 $\Gamma (\mu \to e + \gamma)$ and $\Gamma (\tau \to e +
\gamma)$ decays are still given by the MIAM.

The decay width depends on the SUSY parameters, and again we shall randomly generate
the points and use the current bound $BR(\tau \to \mu+\gamma) <
1.1 \times 10^{-6}$ to determine which percentage is excluded/accepted.
In Fig. 3, we can see  that starting with values of the scalar mass parameter
$\sm0 \geq 460$  GeV, about $100 \%$ of the generated points are acceptable
for $x \geq 0.3$, see Fig. 7 (compare with the result $\sm0 \geq 360$  GeV,
obtained using MIAM).

\section{Radiative Fermion masses in the MSSM}

Understanding the origin of fermion masses and mixing angles is
one of the main problems in Particle Physics. Because of the
observed hierarchy, it is plausible to suspect that some of the
entries in the full (non-diagonal) fermion mass matrices could
be originate as a radiative effect. The MSSM loops involving
sfermions and gauginos some of those entries could generate .
However, most attempts presented so far
\cite{Ferrandis:2004ng,Ferrandis:2004ri,Diaz-Cruz:2005qz,Diaz-Cruz:2000mn}
could be seen as being highly dependent on the details of the SUSY
breaking particular aspects. In this section we would like to scan
the parameter space in order to determine which is the natural size
of such corrections, namely to study which of the fermion masses
could be generated in a natural manner. We shall concentrate on
the charged lepton case, and will use both the MIAM as well as the FDM of a particular {\it
Ansatz} for the soft-breaking trilinear terms.

\subsection{Mass-insertion approximation method (MIAM)}

A Left-Right diagonal mass-insertion
$(\delta_{ii})_{LR}=(\delta_{ii})_{RL}$ generates  a one-loop mass
term for leptons given by \cite{gabbiani}
\beq 
\delta m_{i}=-\frac{\alpha}{2 \pi} \, m_{\tilde{\gamma}} \,
Re(\delta_{ii})_{LR} \, I(x_{\tilde{\gamma}}),
\eeq
where the function $I(x)$ is given by 
\beq I(x) = \frac{-1+x-x \ln(x)}{(1-x)^2}. 
\eeq 
In our approximation 
\beq
Re(\delta_{ii})_{LR} = \frac{\cos\beta}{\sqrt{2}} \frac{v}{\sm0}, \,
(z^l_{LR})_{ii} 
\eeq 
hence 
\beq 
\delta m_{i}=-\frac{\alpha}{2 \pi} \, I(x_{\tilde{\gamma}})
\frac{\cos\beta}{\sqrt{2}} \, \sqrt{x_{\tilde{\gamma}}} \, v
\,(z^l_{LR})_{ii}. \label{eq:delii} 
\eeq 
Again, $ x_{\tilde{\gamma}} \equiv(m_{\tilde{\gamma}}/\sm0)^2$ . Again, we
shall generate $10^5$ random values of $O(1)$ for the parameter
$(z^l_{LR})_{ii}$. In addition such points must satisfy the LFV current
bounds. One can estimate the natural value of the fermion
mass generated from SUSY loops, by taking $x_{\tilde{\gamma}}=0.3$,
$\tan\beta=15-50$ and $(z^l_{LR})_{ii} \approx 1$, which gives
$\delta m_{i} \approx 10-3$ MeV. Thus, in order to generate the
$e$-$\mu$ hierarchy, one will need to include it in the
$A$-terms, namely:
$$\frac{\delta m_e}{\delta m_{\mu}}=\frac{m_e}{m_{\mu}} \cong \frac{1}{200},$$
\noindent then
$$\frac{(z^l_{LR})_{11}}{(z^l_{LR})_{22}} \cong \frac{1}{200}.$$
Such hierarchy  can only arise as a
result of some flavor symmetry. Thus, one can see that radiative
mechanism requires an additional input in order to reproduce the
observed fermion masses. The percentage of
points that produce a correction that falls within the range $0.5 <
\delta m_{e}/m_{e} < 2.0$ as a function of $\tan\beta$, for
$x_{\tilde{\gamma}}=0.1,0.3,1.5,5.0$, is shown in Fig. 5(a); 
the percentage of points that produce a correction that
falls within the range $0.5 < \delta m_{\mu}/m_{\mu} < 2.0$ as a
function of $\tan\beta$, for $x_{\tilde{\gamma}}=0.1$ is plotted in Fig 5(b),
$x_{\tilde{\gamma}}=0.3$, $x_{\tilde{\gamma}}=1.5$ and
$x_{\tilde{\gamma}}=5$. 
We numerically observed  that it is not
possible to generate the tau mass
 (it can be shown that at least one fermion should have a mass in
order to radiatively  generate the rest). Numerically, 
we have found that it is possible to find a set of
parameters $x_{\tilde{\gamma}}$ and $\tan\beta$ for which the
fraction of points that produce a correction that falls
simultaneously within the range $0.5 < \delta m_{e}/m_{e} < 2.0$ and
$0.5 < \delta m_{\mu}/m_{\mu} < 2.0$ is small, but different from
zero, as it is shown in Fig. 6.

It can be noticed that without further theoretical input the values
of $(z^l_{LR})_{ii}$ do not make distinction  between the families. For the electron
mass, one needs higher values of $\tan\beta$ in order to get a significant
fraction of points (bigger than 10\%) where the electron mass is generated.
For lower values of $\tan\beta$, what happen that the mass generated exceeds
the range ($0.5 < \delta m_{e}/m_{e} < 2.0$).

\subsection{Exact diagonalization of a particular {\it Ansatz} and the
one loop correction}

\underline{{{\it 1. Ansatz} A}}

When one uses the exact diagonalization, one can identify the dominant
finite one loop contribution to the lepton mass correction $\delta m_{l}$.
It is given by

\begin{equation}\label{lmasscorr}
(\delta m_{l})_{ab}=
 \frac{\alpha}{2\pi}m_{\tilde{B}}\sum_{c}
 Z_{ca}^{l}Z_{c(b+3)}^{l*}B_{0}
 (m_{\tilde{B}},m_{\tilde{l}_{c}})
\end{equation}

\noindent where $l_{c}$ ($c=4,5,6$)are the lepton left mass eigenstates
($c=1,2,3$) and the lepton right mass eigenstates .
The selectrons can be decoupled with no flavor
mixing with the $\tilde{\mu}-\tilde{\tau}$ sector, then the sfermion
matrix is diagonalized by an unitary matrix, $Z^l$, which is
given on the basis $(\tilde{e}_{L},\tilde{\mu}_{L},\tilde{\tau}_{L},
\tilde{e}_{R},\tilde{\mu}_{R},\tilde{\tau}_{R})$ as follows:
\beq
\bea{l}
\left\lgroup
\bea{l}
\tilde{e}_L\\
\mus_L\\
\tas_L\\
\tilde{e}_R\\
\mus_R\\
\tas_R
\eea
\right\rgroup
\!\!=\!\!
\left\lgroup
\bea{rrrrrr}
  1 & 0 & 0 & 0 & 0 & 0 \\
  0 & c_{1}c_{3} & s_{1}s_{4} & 0 & c_{1}s_{3} & s_{1}c_{4} \\
  0 & -s_{1}c_{3} & c_{1}s_{4} & 0 & -s_{1}s_{3} & c_{1}c_{4} \\
  0 & 0 & 0 & 1 & 0 & 0 \\
  0 & -c_{2}s_{3} & s_{2}c_{4} & 0 & c_{2}c_{3} & -s_{2}s_{4} \\
  0 & s_{2}s_{3} & c_{2}c_{4} & 0 & -s_{2}c_{3} & -c_{2}s_{4}
\eea
\right\rgroup
\!\!
\left\lgroup
\bea{l}
\tilde{e}_1\\
\mus_1\\
\tas_1\\
\tilde{e}_2\\
\mus_2\\
\tas_2
\eea
\right\rgroup , 
\eea
\label{eq:rotationx}
\eeq
with $s_{1,2,3,4}$ defined in Eq. (\ref{eq:rotation2}).

From the rotation matrix (\ref{eq:rotationx}), we see  that matrix
elements $(\delta m_{l})_{a1}=(\delta m_{l})_{1b}=0$, therefore
only the muon mass can be entirely generated  from loop
corrections. The rest of the matrix element are given as follows:
{\small
\begin{equation}
\hspace{-0.4cm}\begin{array}{rcl}
(\delta m_{l})_{22}&=&\frac{\alpha}{4\pi}m_{\tilde{B}}\left\{
[c_{1}^2 B_{0}(m_{\tilde{B}},m_{{\tilde \mu}_{1}})-c_{2}^2
B_{0}(m_{\tilde{B}},m_{{\tilde \mu}_{2}})]+ [s_{1}^2
B_{0}(m_{\tilde{B}},m_{{\tilde \tau}_{1}})-s_{2}^2
B_{0}(m_{\tilde{B}},m_{{\tilde \tau}_{2}})]\right\},
\\
(\delta m_{l})_{23}&=&\frac{\alpha}{4\pi}m_{\tilde{B}} \left\{
c_{1}s_{1} [B_{0}(m_{\tilde{B}},m_{{\tilde
\mu}_{1}})-B_{0}(m_{\tilde{B}},m_{{\tilde \tau}_{1}})]+ c_{2}s_{2}
[B_{0}(m_{\tilde{B}},m_{{\tilde
\mu}_{2}})-B_{0}(m_{\tilde{B}},m_{{\tilde \tau}_{2}})]\right\},
\\
(\delta m_{l})_{32}&=&(\delta m_{l})_{23},\\
(\delta m_{l})_{33}&=&\frac{\alpha}{4\pi}m_{\tilde{B}}\left\{
[s_{1}^2 B_{0}(m_{\tilde{B}},m_{{\tilde \mu}_{1}})-s_{2}^2
B_{0}(m_{\tilde{B}},m_{{\tilde \mu}_{2}})]+[c_{1}^2
B_{0}(m_{\tilde{B}},m_{{\tilde \tau}_{1}})-c_{2}^2
B_{0}(m_{\tilde{B}},m_{{\tilde \tau}_{2}})]\right\},
\end{array}
\end{equation}
}
where
$$B_{0}(m,m_{i})-B_{0}(m,m_{j})=\ln\left(\frac{m_{j}^{2}}{m_{i}^{2}}\right)
+\frac{m^{2}}{m_{i}^{2}-m^{2}}\ln
\left(\frac{m_{i}^{2}}{m^{2}}\right)-\frac{m^{2}}{m_{j}^{2}-m^{2}}\ln
\left(\frac{m_{j}^{2}}{m^{2}}\right)$$
which follows from
$$B_{0}(m_{1},m_{2})= 1+\ln\left(\frac{Q^2}{m_{2}^{2}}\right)+
\frac{m_{1}^{2}}{m_{2}^{2}-m_{1}^{2}}\ln
\left(\frac{m_{2}^{2}}{m_{1}^{2}}\right).$$
After generating $10^5$ random values of $O(1)$ for the parameters
$y$ and $z$, we show our results in Figs. 7 and 8. In Fig. 7 is
shown the percentage of points that produce a correction that falls
within the range $0.5 < \delta m_{\mu}/m_{\mu} < 2.0$ as a function
of $\tan\beta$, for $x_{\tilde{\gamma}}=0.1$,
$x_{\tilde{\gamma}}=0.3$ and $x_{\tilde{\gamma}}=0.5$. We notice
that a high $\tan\beta$ range is required to get a correct
generation. In Fig. 8 is plotted the percentage of points that
produce a correction that falls within the range $0.5 < \delta
m_{\mu}/m_{\mu} < 2.0$ as a function of $\sm0$, for
$x_{\tilde{\gamma}}=0.1$ and $\tan\beta=32$,
$x_{\tilde{\gamma}}=0.3$ and $\tan\beta=56$,
$x_{\tilde{\gamma}}=0.5$ and $\tan\beta=72$. We find that a slepton
mass parameter $\sm0 \lsim 1$ TeV is required in order to generate
the muon mass for about 40-60\% of generated points.

\underline{{{\it Ansatz} B}}

As we have already mentioned in Subsection IV.B, when we use the
exact diagonalization, we can identify the dominant finite one
loop contribution to the lepton mass correction $\delta m_{l}$,
which is given by Eq.(\ref{lmasscorr}). Using the {\it
Ansatz} B (Eq.(\ref{eq:Alepb})), the sfermion matrix is
diagonalized by an unitary matrix, $Z^l_B$, which is given, in the
basis $(\tilde{e}_{L},\tilde{\mu}_{L},\tilde{\tau}_{L},
\tilde{e}_{R},\tilde{\mu}_{R},\tilde{\tau}_{R})$, as:

\beq \bea{l} \left\lgroup \bea{l}
\tilde{e}_L\\
\mus_L\\
\tas_L\\
\tilde{e}_R\\
\mus_R\\
\tas_R \eea \right\rgroup \!\!= \frac{1}{\sqrt{2}} \!\! \left\lgroup
\bea{rrrrrr}
  1 & 0 & 0 & 0 & 0 & 0 \\
  0 & -s_{\xi} & -c_{\xi} & 0 & s_{\xi} & c_{\xi} \\
  0 & c_{\xi} & -s_{\xi} & 0 & -c_{\xi} & s_{\xi} \\
  0 & 0 & 0 & 1 & 0 & 0 \\
  0 & -s_{\xi} & c_{\xi} & 0 & -s_{\xi} & c_{\xi} \\
  0 & c_{\xi} & s_{\xi} & 0 & c_{\xi} & s_{\xi} \\
\eea \right\rgroup \!\! \left\lgroup \bea{l}
\tilde{e}_1\\
\mus_1\\
\tas_1\\
\tilde{e}_2\\
\mus_2\\
\tas_2 \eea
\right\rgroup , 
\eea \label{eq:rotationxb} \eeq

\noindent with $s_{\xi}\equiv \sin(\phi/2)$ and $c_{\xi}\equiv
\cos(\phi/2)$ (see Eq.(\ref{defphi})).

From the rotation matrix (\ref{eq:rotationxb}), we see that matrix
elements $(\delta m_{l})_{a1}=(\delta m_{l})_{1b}=0$, therefore only
the $\mu$ mass can be entirely generated  from loop corrections. The
rest of the matrix element are given as follows:
\begin{eqnarray}
(\delta m_{l})_{22}&=&\frac{\alpha}{4\pi}m_{\tilde{B}}\left\{
s^2_{\xi}[B_{0}(m_{\tilde{B}},m_{{\tilde
\mu}_{2}})-B_{0}(m_{\tilde{B}},m_{{\tilde \mu}_{1}})]
+c^2_{\xi}[B_{0}(m_{\tilde{B}},m_{{\tilde
\tau}_{2}})-B_{0}(m_{\tilde{B}},m_{{\tilde \tau}_{1}})]\right\}
\nonumber\\
(\delta m_{l})_{23}&=&\frac{\alpha}{4\pi}m_{\tilde{B}} \left\{
s_{\xi}c_{\xi} [B_{0}(m_{\tilde{B}},m_{{\tilde
\tau}_{2}})-B_{0}(m_{\tilde{B}},m_{{\tilde \mu}_{2}})]+
s_{\xi}c_{\xi} [B_{0}(m_{\tilde{B}},m_{{\tilde
\tau}_{1}})+B_{0}(m_{\tilde{B}},m_{{\tilde \mu}_{1}})]\right\}
\nonumber\\
(\delta m_{l})_{32}&=&(\delta m_{l})_{23} \\
(\delta m_{l})_{33}&=&\frac{\alpha}{4\pi}m_{\tilde{B}} \left\{
c_{\xi}^2 [B_{0}(m_{\tilde{B}},m_{{\tilde
\mu}_{2}})-B_{0}(m_{\tilde{B}},m_{{\tilde \mu}_{1}})]+ s_{\xi}^2
[B_{0}(m_{\tilde{B}},m_{{\tilde
\tau}_{2}})-B_{0}(m_{\tilde{B}},m_{{\tilde \tau}_{1}})]\right\} \, ,
\nonumber
\end{eqnarray}
where $B_{0}(m,m_{i})-B_{0}(m,m_{j})$ and $B_{0}(m_{1},m_{2})$ are
given in the previous Subsection (IV.B).

After generating $10^5$ random values of $O(1)$ for the parameters
$w$ and $y$, we show our results in Figs. 9 and 10. In Fig. 9 is
shown the percentage of points that produce a correction that falls
within the range $0.5 < \delta m_{\mu}/m_{\mu} < 2.0$ as a function
of $\tan\beta$, for $x_{\tilde{\gamma}}=0.05$,
$x_{\tilde{\gamma}}=0.1$ and $x_{\tilde{\gamma}}=0.2$. We notice
that a $0 \lsim \tan\beta \lsim 10$ range is required to get a
correct generation. The percentage of
points that produce a correction that falls within the range $0.5 <
\delta m_{\mu}/m_{\mu} < 2.0$ as a function of $\sm0$, for
$x_{\tilde{\gamma}}=0.05$ and $\tan\beta=3.2$,
$x_{\tilde{\gamma}}=0.1$ and $\tan\beta=4.2$,
$x_{\tilde{\gamma}}=0.2$ and $\tan\beta=4.7$, is plotted in Fig. 10. We found that a slepton
mass parameter $\sm0 \lsim 1$ TeV is required in order to generate
the muon mass for about 40-65\% of the generated points.

\section{Conclusions}

We have discussed
the SUSY flavor problem in the lepton sector using the mass-insertion
approximation, evaluating the radiative LFV loop transitions
($l_i \to l_j \gamma$) with a random generation of the slepton $A$-terms.
Our results illustrate the severity of the SUSY flavor problem for low
sfermion masses. One can see that even for $\sm0=1$ TeV almost $100 \%$ of
the randomly generated points are excluded, while one needs
to have $\sm0 \approx 10$ TeV in order to get about $10 \%$ of the
generated points that satisfy the current bound on $\mu \to e \, \gamma$, having larger
gaugino helps to ameliorate the problem, but not by much. On the other hand,
we have shown that current bounds on tau decays pose no such a severe
problem. In this case, most of the randomly generated points satisfy the
experimental bounds on $\tau \to \mu \, \gamma$ and $\tau \to e \, \gamma$.
Also, we presented two {\it Ansaetze} for soft breaking trilinear terms, the
diagonalization of the resulting sfermion mass matrices, and repeat
the previous calculation.  We showed that for $\sm0 \geq 460$ GeV, $100 \%$ of
the points are acceptable for $x_{\tilde{\gamma}} \geq 0.3$, with similar behavior in both cases (to be
compared with $\sm0 \geq 360$ GeV obtained using the mass-insertion
approximation).

The radiative generation of fermion masses within the context of the MSSM with general trilinear
soft-breaking terms was discussed in detail. 
We presented results for
slepton spectra for $\sm0=100,500$ GeV and $\sm0=1,10$ TeV, with $\tan\beta=15$,
showing that both $\tas_1$ and $\tas_2$ differ significantly from $\sm0$. Moreover, $\tas_1$ can
be as light as $100-300$ GeV, which will have an important
effect in the loop calculations. Furthermore, $m_{\mus_i}$ can differ from $\sm0$ for 30-50
GeV considering $z \simeq
0.5$; with these mass values the slepton phenomenology would have
to be reconsidered. 
We also observed that $m_{\mus_1} - m_{\tas_1}$
and $m_{\mus_2} - m_{\tas_2}$  almost behave constant as one varies
the parameter $z$ in the range
$0 \leq z \leq 1$. This splitting affects LFV transitions and radiative fermion mass
generation results. 

Also, we have analyzed the radiative generation of the $e$ and $\mu$
masses using the MIAM by generating $10^5$ random
values of $O(1)$ for the parameters $(z^l_{LR})_{ii}$.
It was shown that for some parameters a percentage of points may produce
a correction that falls within the range $0.5 < \delta m_{l}/m_{e} < 2$,
while another percentage of points can produce a correction that falls
within the range $0.5 < \delta m_{l}/m_{\mu} < 2$. Then,
it is possible to find a set of parameters $x$ and
$\tan\beta$ for which the fraction of points that produce a correction that
falls simultaneously within the range $0.5 < \delta m_{e}/m_{e},\delta m_{\mu}/m_{\mu} < 2.0$, which is small, 
but different from zero. Numerically concluding that it is not possible to generate the tau mass.
Having noticed that without further theoretical input the values
of $(z^l_{LR})_{ii}$ do not distinguish among the families. For the electron
mass, one needs higher values of $\tan\beta$ in order to get a significant
fraction of points (bigger than 10\%) where the electron mass is generated.
For lower values of $\tan\beta$, what happens is that the mass generated exceeds
the range ($0.5 < \delta m_{e}/m_{e} < 2.0$).

We have pointed out that in order to generate the $e$-$\mu$ hierarchy,
one needs to have $(z^l_{LR})_{11}/(z^l_{LR})_{22} \cong 1/200$.
Such hierarchy can only arise as a result of
some flavor symmetry. Thus, one can conclude that the radiative mechanism
requires an additional input in order to reproduce the observed fermion
masses.

On the other hand, we have analyzed the radiative generation of
the muon mass using a FDM, by considering
at the weak scale two different {\it Ansaetze} for 
$A$-term, by generating $10^5$
random values of $O(1)$ for the parameters $y$ and $z$ of the
model. It is shown that for some parameters a percentage of points
may produce a correction that falls within the range $0.5 < \delta
m_{\mu}/m_{\mu} < 2$, watching a quite different behavior from the resulting
fractions of acceptable points when we
consider the different {\it Ansaetze} as well as with the two full
diagonalization models and the mass-insertion approximation. Similarly to the mass-insertion
approximation case, it is not numerically
possible to radiatively generate  the tau mass by using the two
full diagonalization models considered.

\bigskip

\noindent{\bf {\large Acknowledgments}}

\noindent We would like to thank C.P. Yuan and H.J. He for
valuable discussions. This work was supported in part by CONACYT
and SNI (M\'exico).

\newpage

\begin{figure}
\includegraphics[width=2.5in,angle=-90]{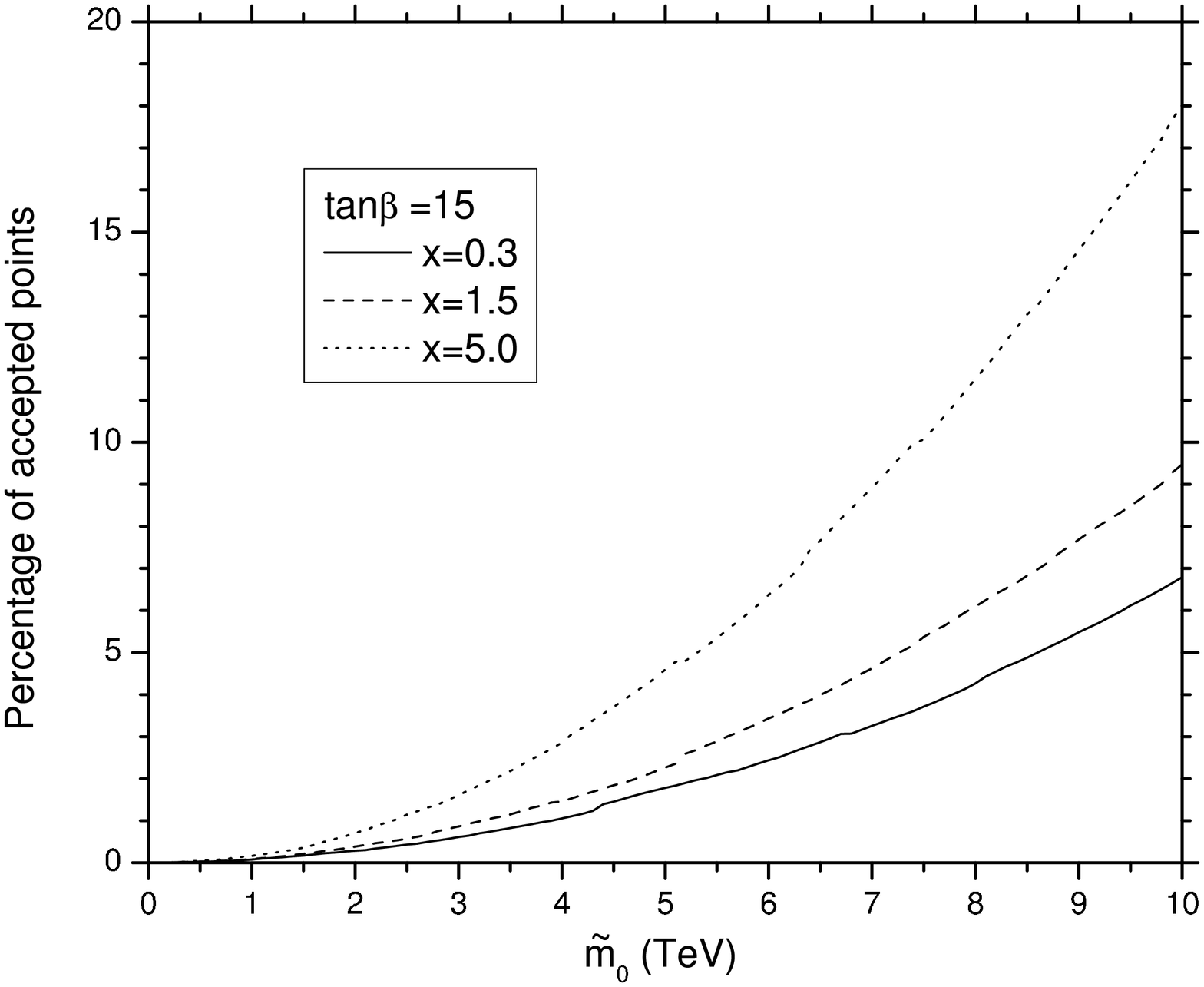}
\vspace{0.5in}
\caption{Analysis of the LFV decay $\mu \to e \, \gamma$ as a
function of $\sm0$, using the MIAM and by
randomly generating  $10^5$ points for  $(z^l_{LR})_{21}$
coefficient, assuming $\tan\beta=15$ for
$x_{\bino}=0.3,\,1.5,\,5$. The different draw-lines show the
fraction of such points that satisfies the current experimental
bound $BR(\mu \to e \, \gamma) < 1.2 \times 10^{-11}$.}
\label{figure1}
\end{figure}

\begin{figure}[floatfix]
\begin{center}
\includegraphics[width=2.3in,angle=-90]{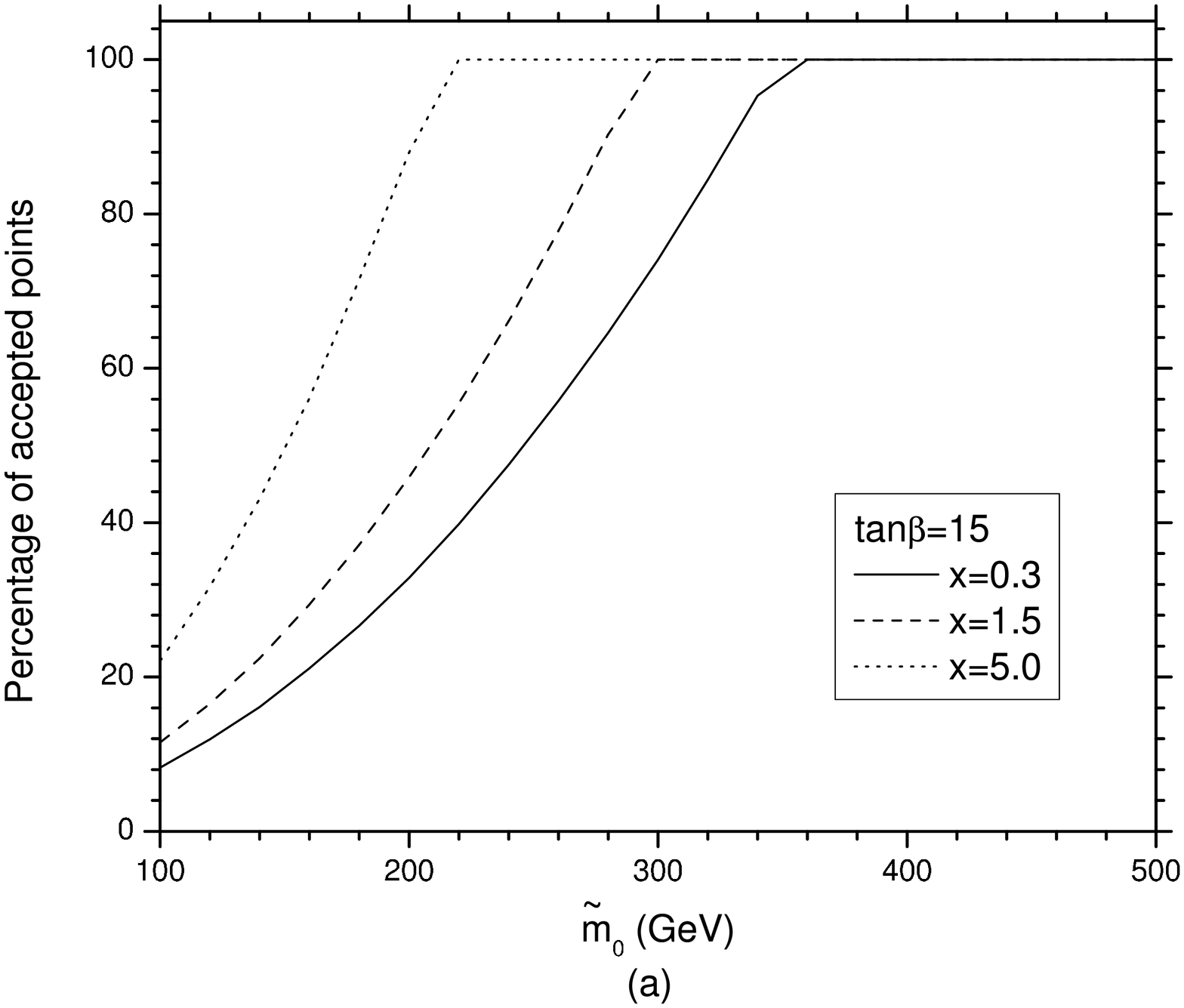}
\hspace{0.7in}
\includegraphics[width=2.3in,angle=-90]{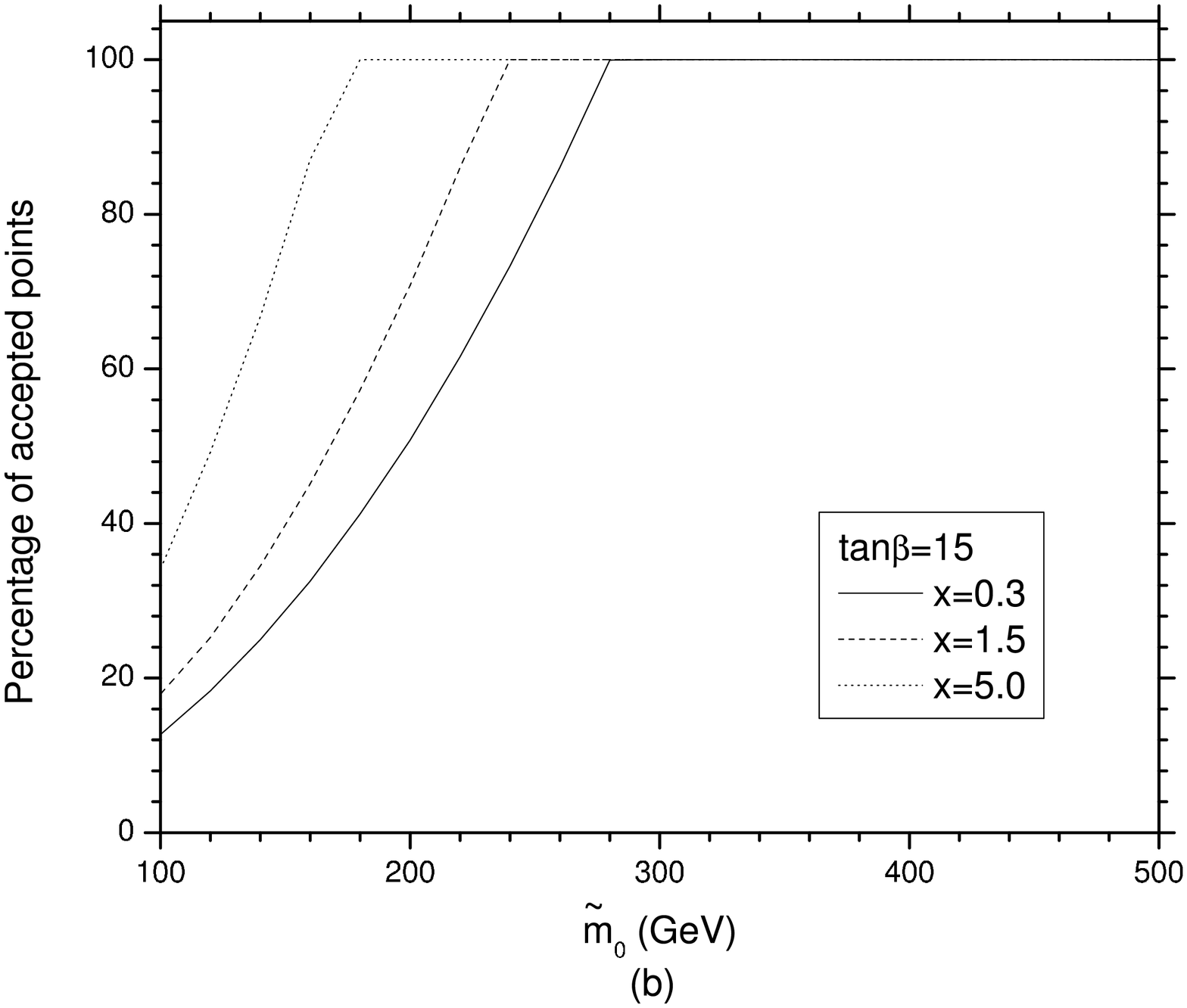}
\end{center}
\vspace{0.5in}
\caption{Analysis of the LFV decays $\tau \to \mu \, \gamma$ and $\tau \to e \, \gamma$ as a
function of $\sm0$, using MIAM and by
randomly generating  $10^5$ points for (a)$(z^l_{LR})_{32}$ and (b) $(z^l_{LR})_{31}$
coefficients, assuming $\tan\beta=15$ and
$x_{\bino}=0.3,\,1.5,\,5$. The different draw-lines show the
fraction of such points that satisfies the current experimental
bounds (a) $BR(\tau \to \mu \, \gamma) < 1.1 \times 10^{-6}$ and (b) $BR(\tau \to e \, \gamma) < 2.7 \times 10^{-6}$.}
\label{figure2}
\end{figure}


\begin{figure}[floatfix]
\includegraphics[width=2.3in,angle=-90]{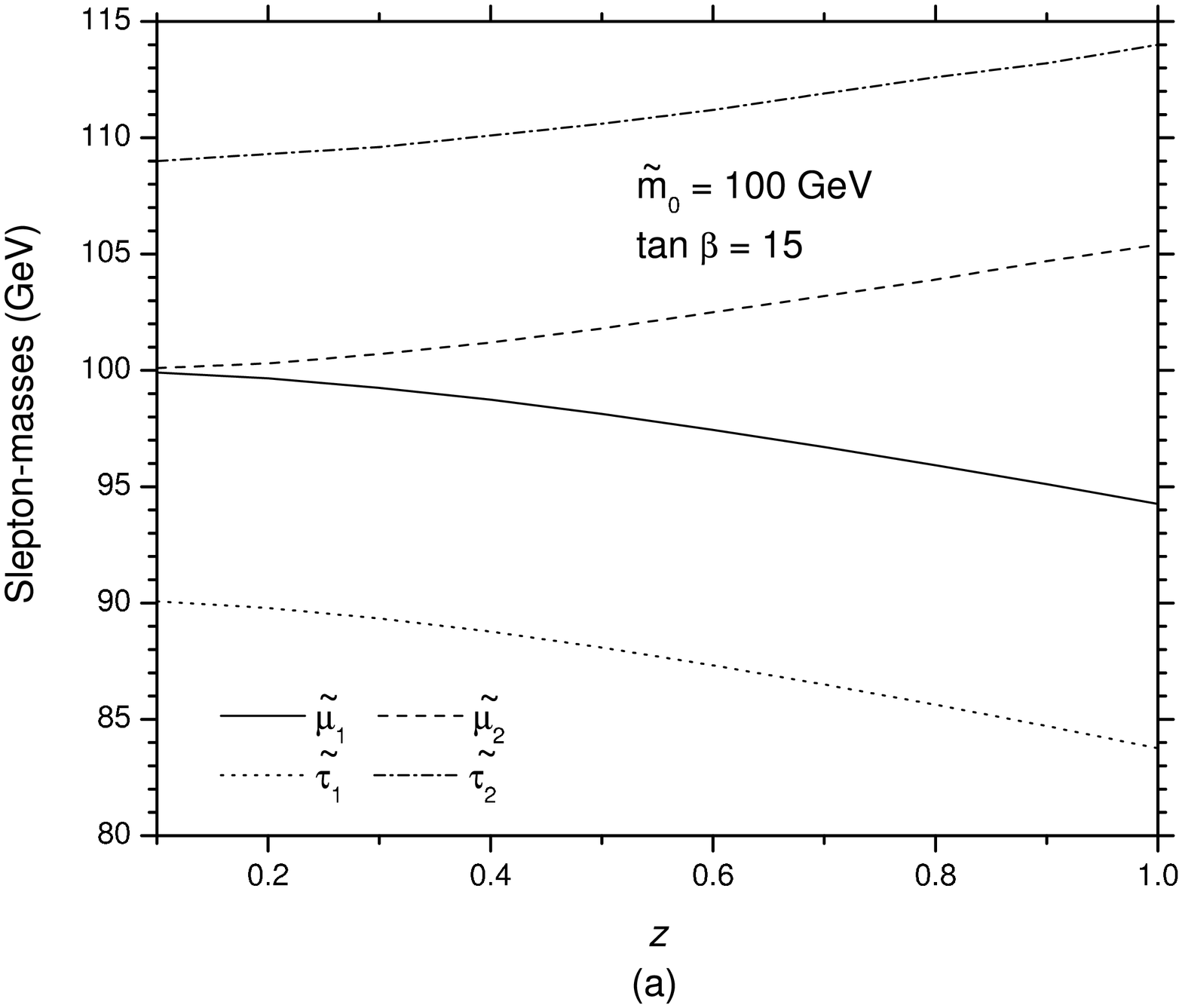}
\hspace{0.7in}
\includegraphics[width=2.3in,angle=-90]{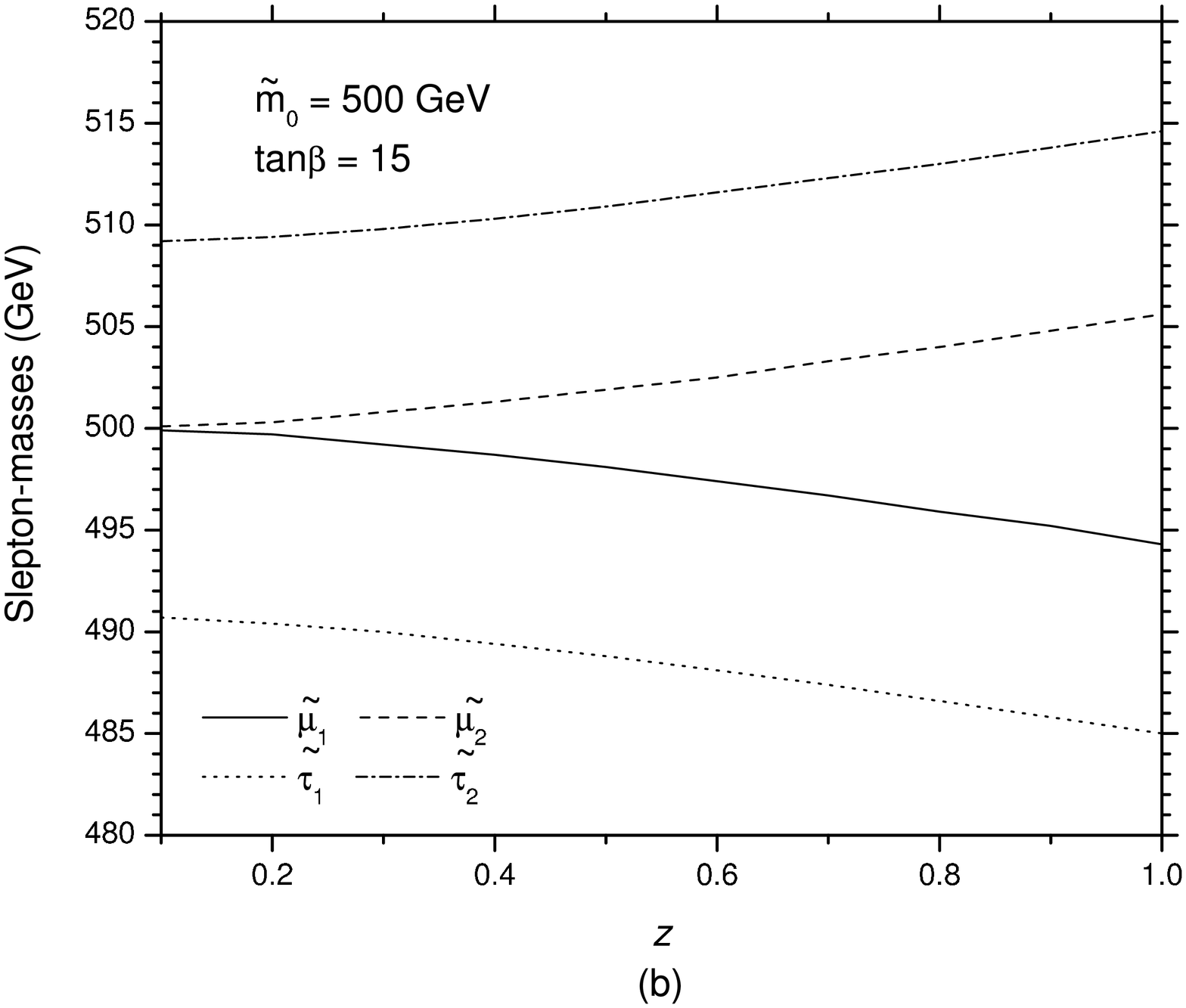}

\vspace{0.5in} 

\includegraphics[width=2.3in,angle=-90]{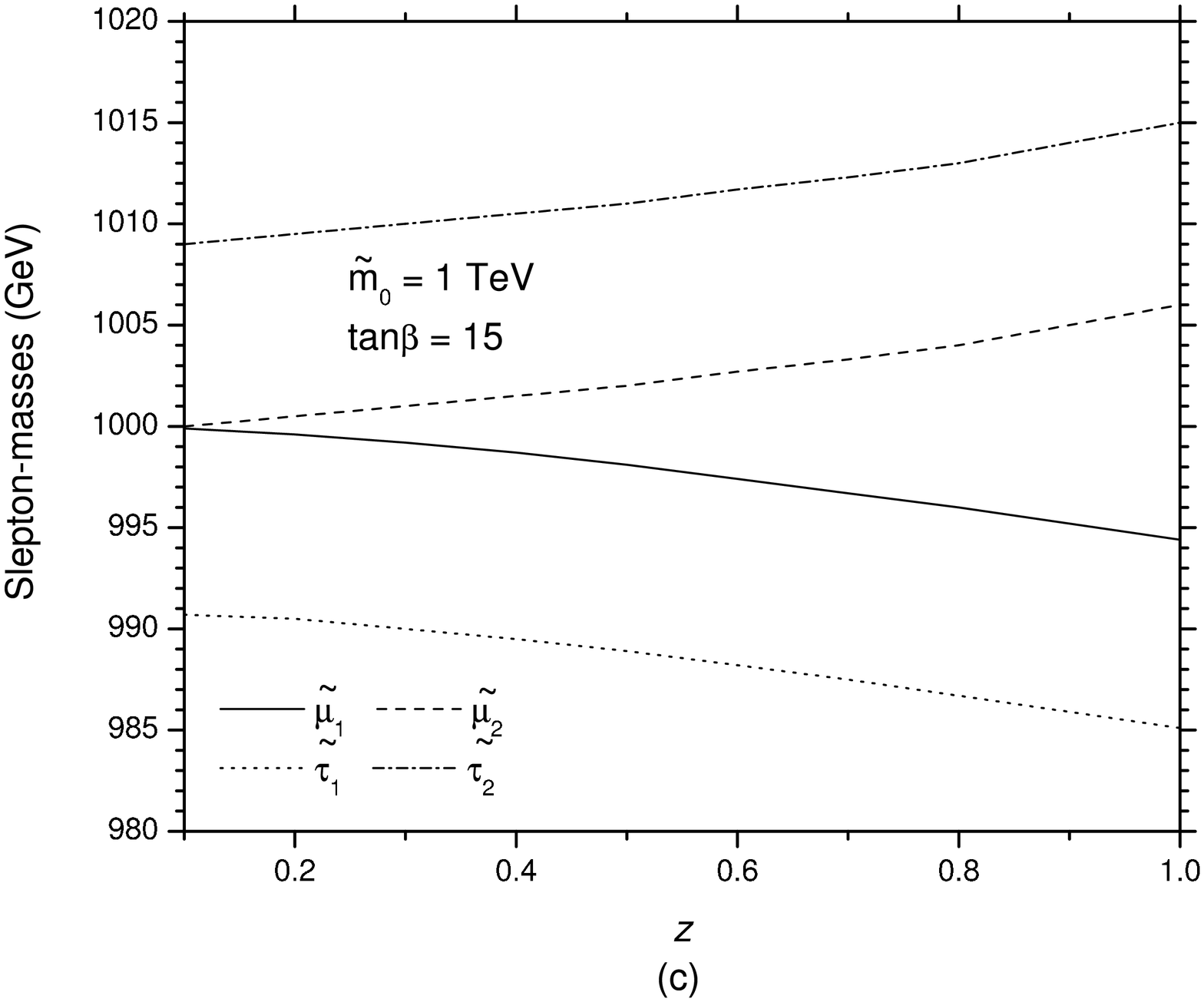}
\hspace{0.7in}
\includegraphics[width=2.3in,angle=-90]{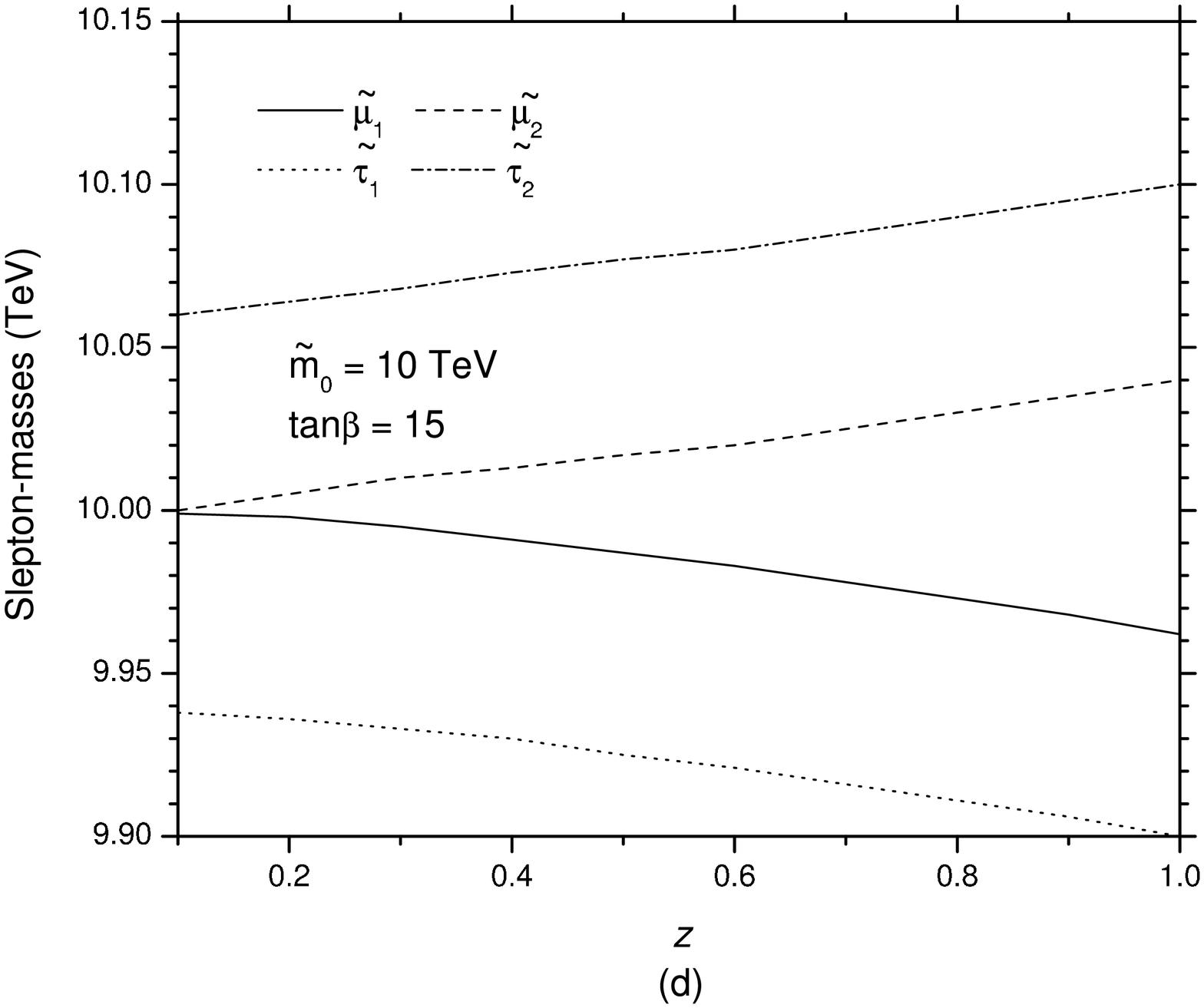}
\vspace{0.5in}
\caption{Mass spectrum for the smuon and stau sleptons as a
function of $z$ for $\tanB=15$ and the SUSY scale (a) $\sm0=100$ GeV, (b) $\sm0=500$ GeV, (c) $\sm0=1$ TeV and  (d) $\sm0=10$ TeV.}
\label{figure4}
\end{figure}
%
%
%

\begin{figure}
\hspace{-2.0in}
\begin{center}
\includegraphics[width=2.5in,angle=-90]{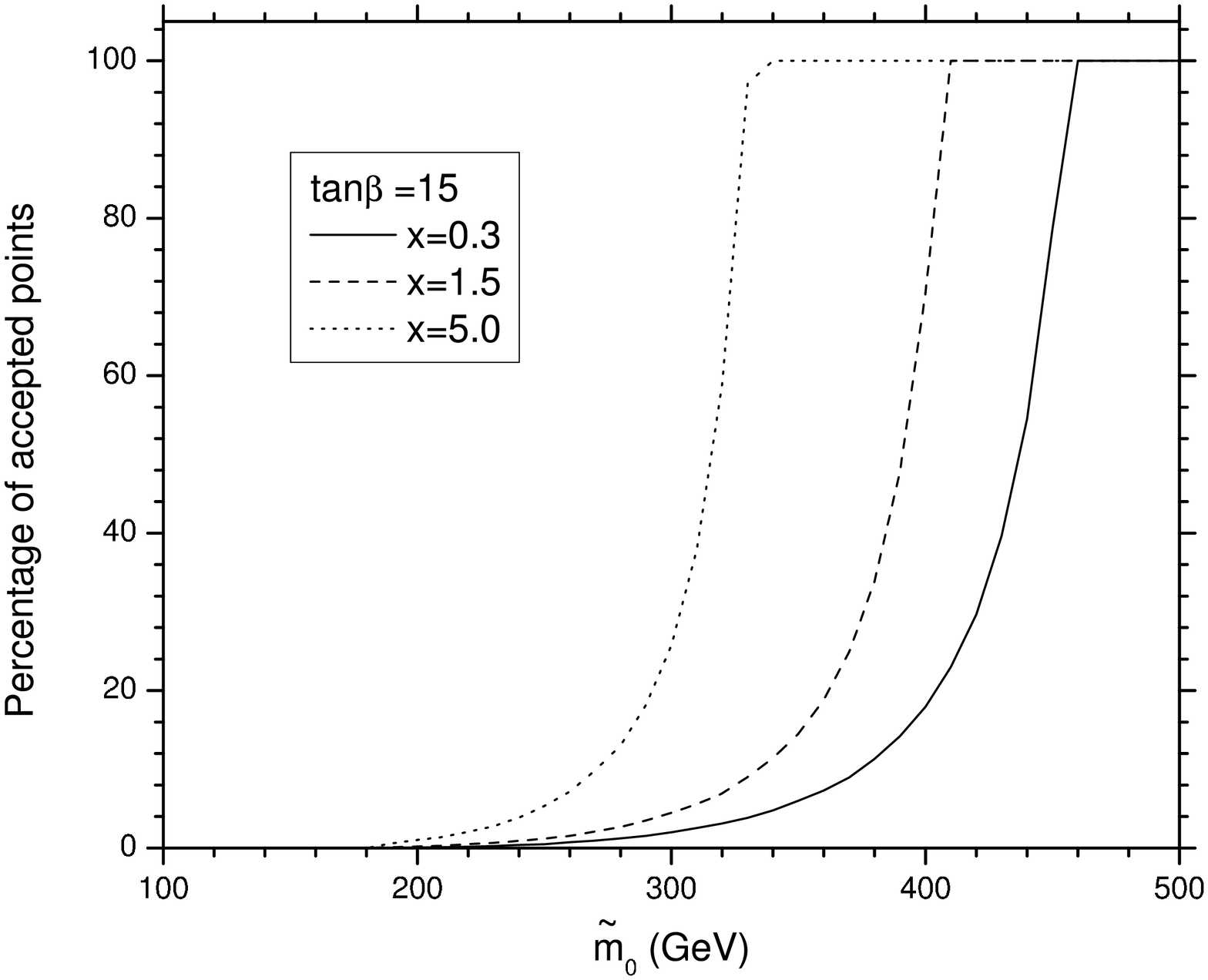}
\end{center}
\vspace{0.6in}
\caption{Analysis of the LFV decay $\tau \to \mu \, \gamma$ as a
function of $\sm0$, using a FDM and by
randomly generating  $10^5$ points for $z$ coefficient,
assuming $\tan\beta=15$ and $x_{\bino}=0.3,\,1.5,\,5$. The
different draw-lines show the fraction of such points that
satisfies the current experimental bound $BR(\tau \to \mu \,
\gamma) < 1.1 \times 10^{-6}$.} \label{figure8}
\end{figure}

\begin{figure}[floatfix]
\begin{center}
\includegraphics[width=2.4in,angle=-90]{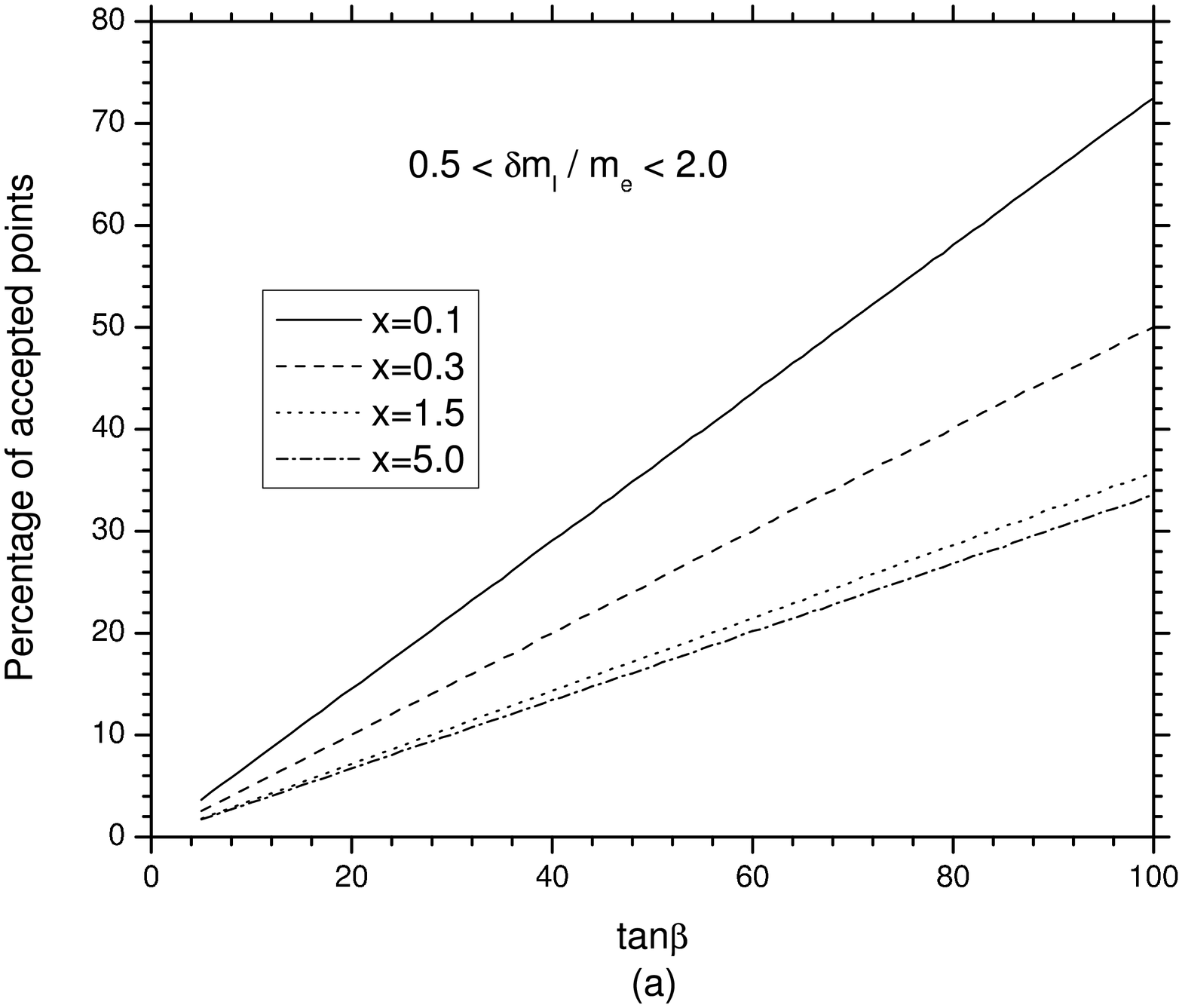}
\hspace{0.7in}
\includegraphics[width=2.4in,angle=-90]{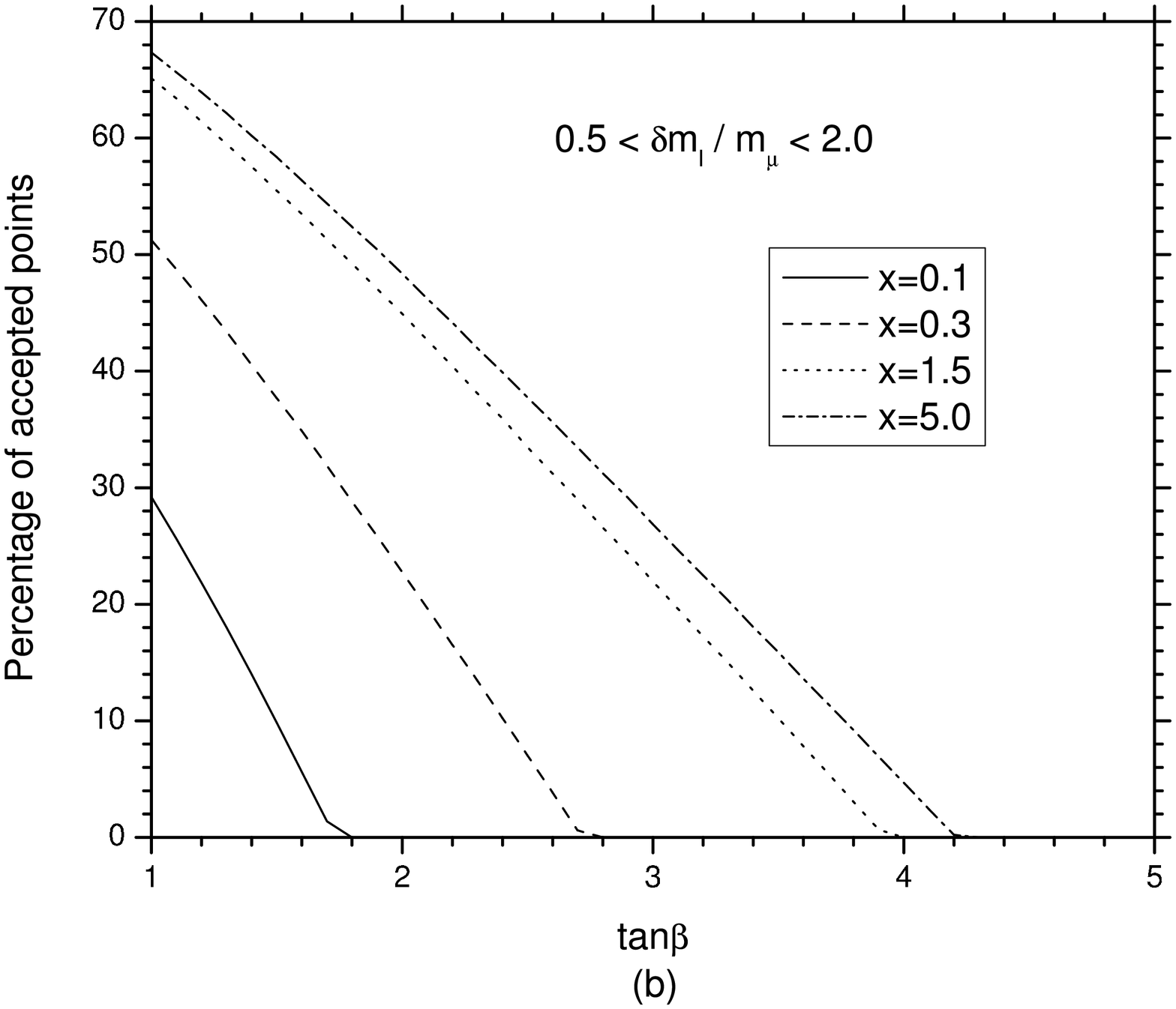}
\vspace{0.5in}
\caption{Radiative generation of the $m_{e}$ and $m_{\mu}$  as a function of
$\tan\beta$, using MIAM and by
generating $10^5$ random values for (a) $(z^l_{LR})_{11}$ and (b) $(z^l_{LR})_{22}$, with $x_{\tilde{\gamma}}=0.1,\,0.3,\,1.5,\,5$. The
different draw-lines show the fraction of points that produce a
correction that falls within the range $0.5 < \delta m_{l}/m_{e},< \delta m_{l}/m_{\mu} <
2.0$.} \label{figure9}
\end{center}
\end{figure}
%

\begin{figure}[floatfix]
\begin{center}
\includegraphics[width=2.5in,angle=-90]{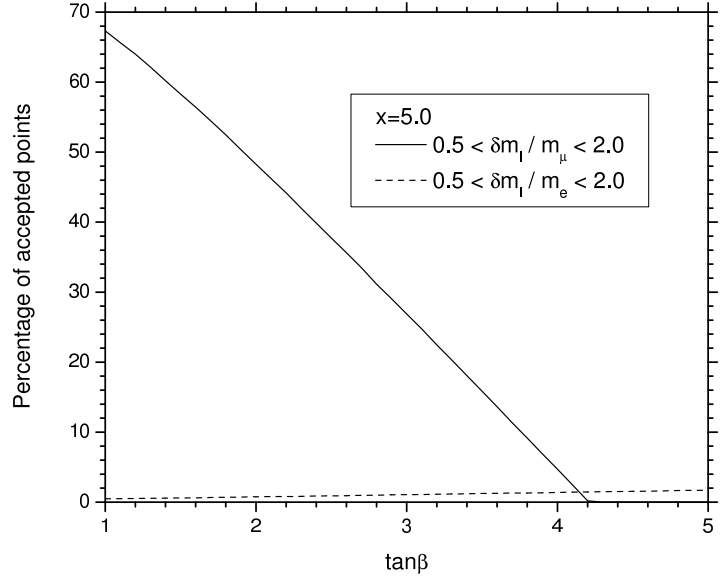}
\vspace{0.5in}
\caption{Radiative generation of the $m_{\mu}$ and $m_{e}$  as a
function of $\tan\beta$, using MIAM and
by generating $10^5$ random values for
$(z^l_{LR})_{22}=(z^l_{LR})_{11}$ with $x_{\tilde{\gamma}}=5$. The
solid draw-line shows the fraction of points that produce a
correction that falls within the range $0.5 < \delta m_{l}/m_{\mu} <
2.0$, while the dashed one shows the fraction of points that produce
a correction that falls within the range $0.5 < \delta m_{l}/m_{e} <
2.0$.} \label{figure11}
\end{center}
\end{figure}

\begin{figure}[floatfix]
\begin{center}
\includegraphics[width=2.5in,angle=-90]{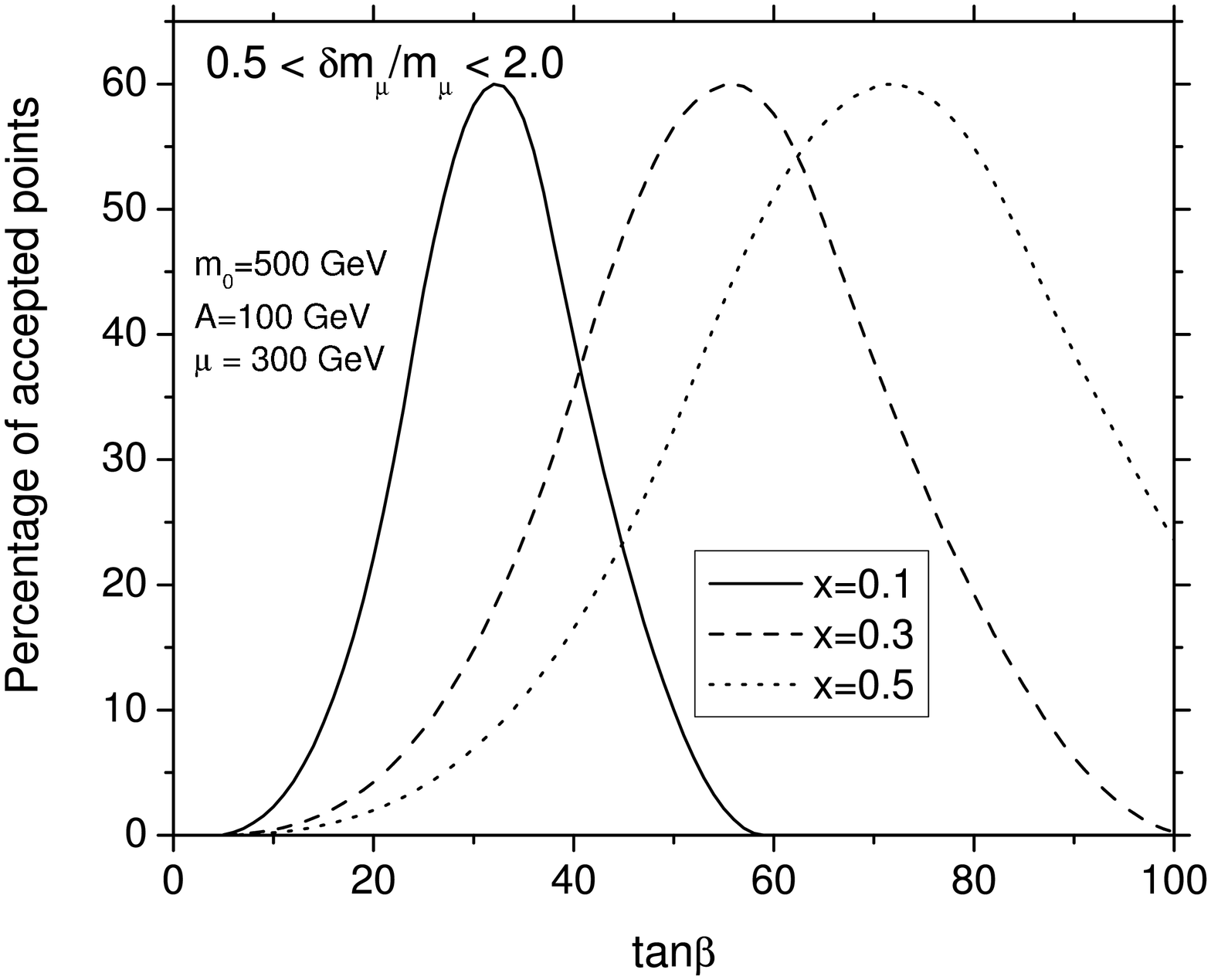}
\vspace{0.5in}
\caption{Radiative generation of the $m_{\mu}$ as a function of
$\tan\beta$, using the FDM A and by
generating $10^5$ random values for $y$ and
$z$ with $x_{\tilde{\gamma}}=0.1,\,0.3,\,0.5$. The different
draw-lines show the fraction of points that produce a correction
that falls within the range $0.5 < \delta m_{\mu}/m_{\mu} < 2.0$.}
\label{figure12}
\end{center}
\end{figure}

\begin{figure}[floatfix]
\begin{center}
\includegraphics[width=2.5in,angle=-90]{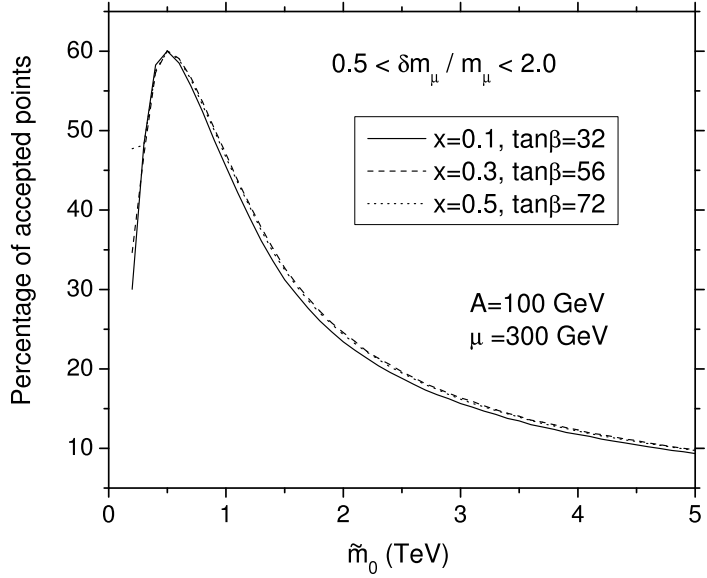}
\vspace{0.5in}
\caption{Radiative generation of the muon mass as a function of
$\sm0$, using the FDM A and by generating
$10^5$ random values  for $y$ and $z$, with:
a) $x_{\tilde{\gamma}}=0.1$ and $\tan\beta=32$, b)
$x_{\tilde{\gamma}}=0.3$ and $\tan\beta=56$, c)
$x_{\tilde{\gamma}}=0.5$ and $\tan\beta=72$. The different
draw-lines show the fraction of points that produce a correction
that falls within the range $0.5 < \delta m_{\mu}/m_{\mu} < 2.0$.}
\label{figure13}
\end{center}
\end{figure}

\begin{figure}[floatfix]
\begin{center}
\includegraphics[width=2.5in,angle=-90]{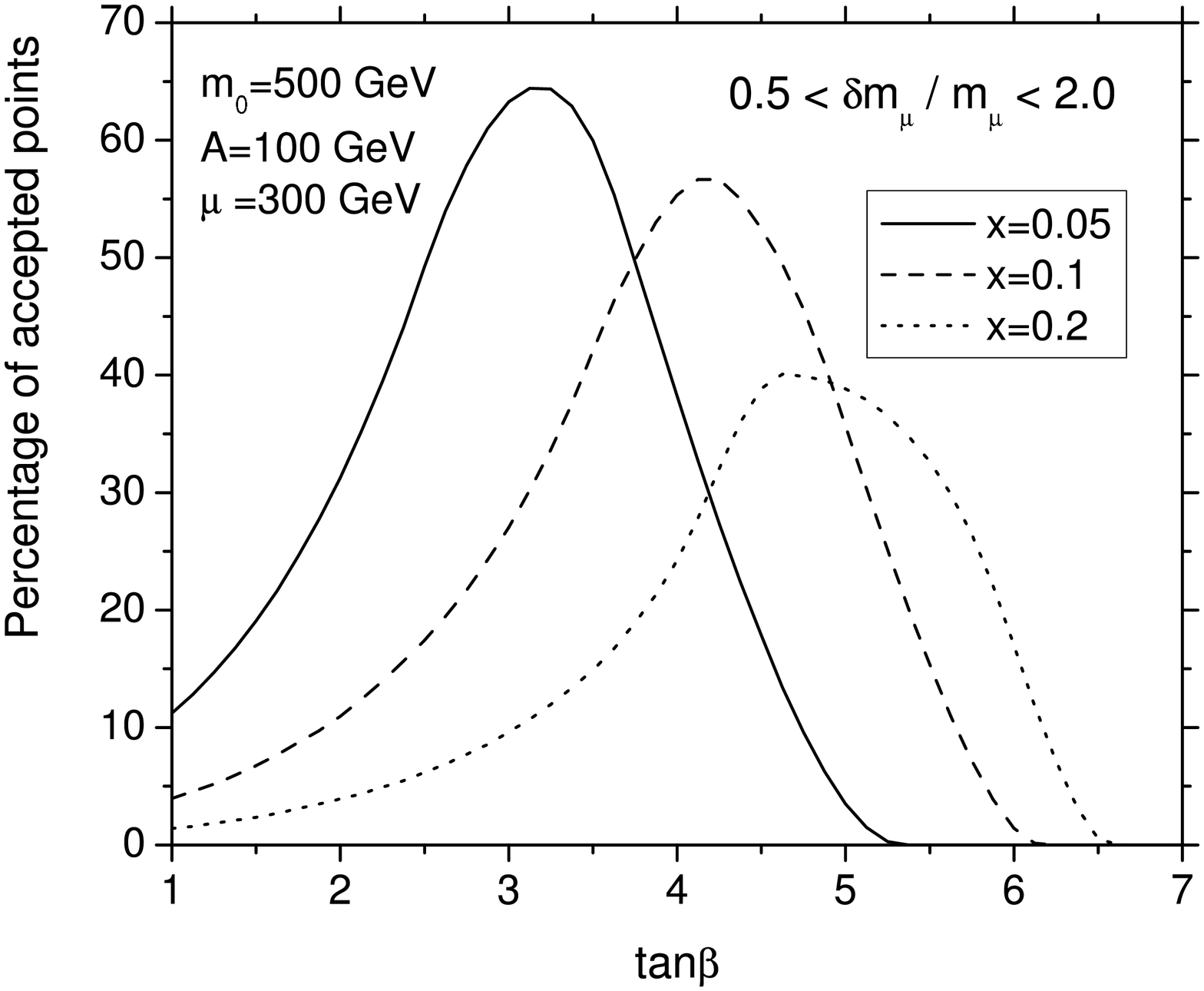}
\vspace{0.5in}
\caption{Radiative generation of the muon mass as a function of
$\tan\beta$, using the FDM B and by
generating $10^5$ random values  for $w$ and
$y$, with $x_{\tilde{\gamma}}=0.05,\,0.1,\,0.2$. The different
draw-lines show the fraction of points that produce a correction
that falls within the range $0.5 < \delta m_{\mu}/m_{\mu} < 2.0$.}
\label{figure14}
\end{center}
\end{figure}

\begin{figure}[floatfix]
\begin{center}
\includegraphics[width=2.5in,angle=-90]{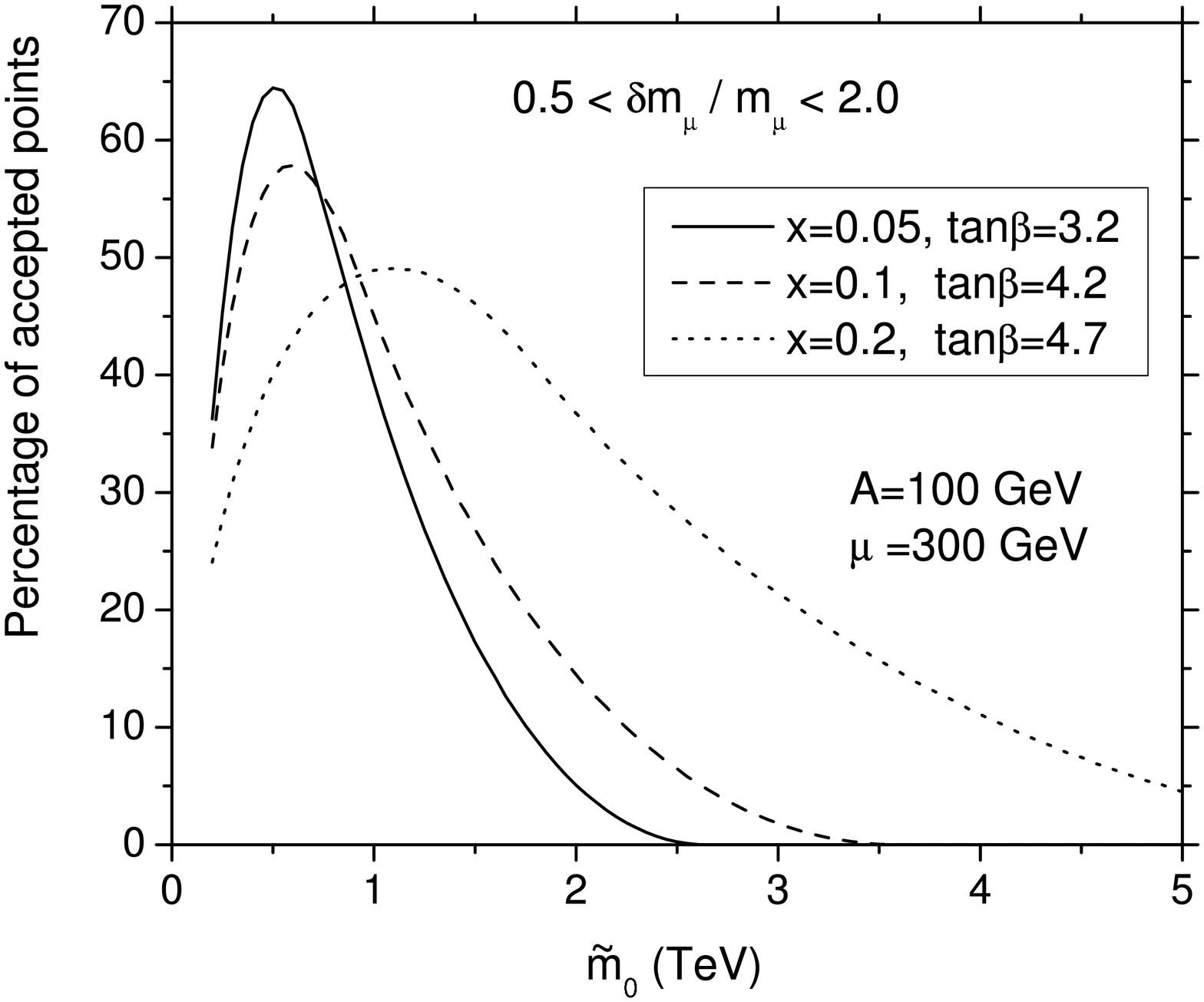}
\vspace{0.5in}
\caption{Radiative generation of the muon mass as a function of
$\sm0$, using the FDM B and by generating
$10^5$ random values for $w$ and $y$, with:
a) $x_{\tilde{\gamma}}=0.05$, $\tan\beta=3.2$, b)
$x_{\tilde{\gamma}}=0.1$, $\tan\beta=4.2$  and c)
$x_{\tilde{\gamma}}=0.2$, $\tan\beta=4.7$. The different
draw-lines show the fraction of points that produce a correction
that falls within the range $0.5 < \delta m_{\mu}/m_{\mu} < 2.0$.}
\label{figure15}
\end{center}
\end{figure}

\begin{references}
\bibitem{review}
See, for instance, recent reviews in ``Perspectives on Supersymmetry'',
ed. G.\,L. Kane, World Scientific Publishing Co., 1998;
H.\,E. Haber, Nucl. Phys. Proc. Suppl. {\bf 101}, 217 (2001),
[hep-ph/0103095].

\bibitem{Chung:2003fi}
D.~J.~H.~Chung, L.~L.~Everett, G.~L.~Kane, S.~F.~King, J.~D.~Lykken and L.~T.~Wang,
Phys.\ Rept.\  {\bf 407}, 1 (2005), [hep-ph/0312378].

\bibitem{nondA}
E.g.,
S. Khalil, J. Phys. G{\bf 27}, 1183 (2001), [hep-ph/0011330];
D. F. Carvalho, M. E. Gomez, S. Khalil, [hep-ph/0104292];
and references therein.

\bibitem{Mura}
A. Masiero and H. Murayama,
Phys. Rev. Lett. {\bf 83}, 907 (1999),
[hep-ph/9903363].

\bibitem{gabbiani}
F. Gabbiani, E. Gabrielli, A. Masiero and L. Silvestrini,
Nucl. Phys. B {\bf 477}, 321 (1996).

\bibitem{Arkani-Hamed:1997ab}
  N.~Arkani-Hamed and H.~Murayama,
  Phys.\ Rev.\ D {\bf 56}, 6733 (1997),
  [hep-ph/9703259].

\bibitem{seiberg}
E.g., Y. Nir and N. Seiberg,
Phys. Lett. B {\bf 309}, 337 (1993),
[hep-ph/9304307].

\bibitem{FCNC}
For review, M. Misiak, S. Pokorski, J. Rosiek,
``Supersymmetry and FCNC Effects'',
[hep-ph/9703442],
in {\it Heavy Flavor II,}  pp.\,795,
Eds. A. J. Buras and M. Lindner,
Advanced Series on Directions in High Energey Physics,
World Scientific Publishing Co., 1998,
and references therein.

\bibitem{oursqmix}
J.~L.~Diaz-Cruz, H.~J.~He and C.~P.~Yuan,
``Soft SUSY breaking, stop-scharm mixing and Higgs signatures,''
Phys.\ Lett.\ B {\bf 530}, 179 (2002),
[hep-ph/0103178].

\bibitem{superkam} Super-Kamiokande Collaboration
(Y. Fukuda et al.), Phys. Rev. Lett. {\bf 81}, 1562 (1998), 
[hep-ex/0009001].

\bibitem{Paradisi:2005fk}
  P.~Paradisi,
  JHEP {\bf 0510}, 006 (2005), 
  [hep-ph/0505046].

\bibitem{Diaz-Cruz:2002er}
  J.~L.~Diaz-Cruz,
  JHEP {\bf 0305}, 036 (2003),
  [hep-ph/0207030].

\bibitem{Diaz-Cruz:1999xe}
  J.~L.~Diaz-Cruz and J.~J.~Toscano,
  Phys.\ Rev.\ D {\bf 62}, 116005 (2000), 
  [hep-ph/9910233].

\bibitem{CCBVS}
J. A. Casas and S. Dimopolous,
Phys. Lett. B {\bf 387}, 107 (1996),
[hep-ph/9606237].

\bibitem{partdata} S. Eidelman {\it et al.}, ({\it Review of Particle
Physics}), Phys. Lett. B {\bf 592}, 1 (2004).

\bibitem{hisanoetal}
J.~Hisano, T.~Moroi, K.~Tobe and M.~Yamaguchi,
``Lepton-Flavor Violation via Right-Handed Neutrino Yukawa Couplings
in Supersymmetric Standard Model,''
Phys.\ Rev.\ D {\bf 53}, 2442 (1996),
[hep-ph/9510309].

\bibitem{Ferrandis:2004ng}
  J.~Ferrandis,
  Phys.\ Rev.\ D {\bf 70}, 055002 (2004),
  [hep-ph/0404068].

\bibitem{Ferrandis:2004ri}
  J.~Ferrandis and N.~Haba,
  Phys.\ Rev.\ D {\bf 70}, 055003 (2004),
  [hep-ph/0404077].

\bibitem{Diaz-Cruz:2005qz}
  J.~L.~Diaz-Cruz and J.~Ferrandis,
  Phys.\ Rev.\ D {\bf 72}, 035003 (2005),
  [hep-ph/0504094].

\bibitem{Diaz-Cruz:2000mn}
  J.~L.~Diaz-Cruz, H.~Murayama and A.~Pierce,
  Phys.\ Rev.\ D {\bf 65}, 075011 (2002),
  [hep-ph/0012275].
\end{references}
\end{document}